\documentclass[10pt,oneside]{article}
\usepackage{graphics}
\usepackage{latexsym}
\usepackage{amsmath}
\usepackage{amssymb}
\usepackage{enumerate}
\usepackage{mathrsfs}
\usepackage[dvips]{graphicx}
\usepackage{psfrag}
\usepackage{caption}
\usepackage{subfigure}

\newtheorem{lem}{Lemma}
\newtheorem{coro}{Corollary}
\newtheorem{prop}{Proposition}

\newcommand{\cP}{\mathcal{P}}

\renewcommand{\theequation}{\thesection.\arabic{equation}}
\numberwithin{equation}{section}

\title{Propagation of Slepyan's crack in a non-uniform elastic lattice}
\author{M.J. Nieves$^\dagger$, A.B. Movchan\footnote{Department of Mathematical Sciences, University of Liverpool, Liverpool L69 3BX, U.K.}, I.S. Jones\footnote{School of Engineering, John Moores University, James Parsons Building, Byrom Street, Liverpool L3 3AF, U.K.} 
   \, and G.S. Mishuris\footnote{Institute of Mathematics and Physics, Aberystwyth University, Aberystwyth, SY23 3BZ.$\>\>\>\>\>\>\>\>\>\>\>\>\>\>\>\>\>\>\>\>\>\>\>\>\>\>\>\>\>\>\>\>\>\>\>\>\>\>\>\>\>\>\>\>\>\>\>\>\>\>\>\>\>\>\>\>\>\>\>\>\>\>\>\>\>\>\>\>\>\>\>\>\>\>\>\>\>\>\>\>\>\>\>\>$ {Corresponding author: M.J. Nieves, Email: M.J.Nieves@ljmu.ac.uk, Tel: +44 151 231 2253} }}
\date{}

\newcommand{\bfm}[1]{\mbox{\boldmath ${#1}$}}

\newcommand{\beqa}{\begin{eqnarray}}
\newcommand{\eeqa}{\end{eqnarray}}
\newcommand{\bequ}{\begin{equation}}
\newcommand{\eequ}[1]{\label{#1}\end{equation}}

\newcommand{\GX}{\Xi}

\newcommand{\BGO}{\bfm\Omega}
\newcommand{\BGX}{\bfm\Xi}

\newcommand{\CA}{{\cal A}}
\newcommand{\CB}{{\cal B}}
\newcommand{\CC}{{\cal C}}

\newcommand{\CJ}{{\cal J}}

\newcommand{\CP}{{\cal P}}

\newcommand{\CZ}{{\cal Z}}

\def\Be{{\bf e}}

\def\Bu{{\bf u}}
\def\Bv{{\bf v}}

\def\Bx{{\bf x}}

\def\BB{{\bf B}}
\def\BC{{\bf C}}

\def\BR{{\bf R}}

\def\BT{{\bf T}}

\newcommand{\beq}{\begin{equation}}
\newcommand{\eeq}{\end{equation}}
\newcommand{\overliner}{\begin{eqnarray}}
\newcommand{\earr}{\end{eqnarray}}
\newcommand{\beqn}{\begin{equation*}}
\newcommand{\eeqn}{\end{equation*}}
\newcommand{\overlinern}{\begin{eqnarray*}}
\newcommand{\earrn}{\end{eqnarray*}}

\newcommand{\sign}{\mbox{sign}}

\begin{document}
\maketitle

\begin{abstract}
We model and derive the solution for the problem of  a Mode I semi-infinite crack propagating  
in a  discrete triangular lattice with  bonds having a contrast in stiffness in the principal lattice directions. 
The corresponding Green's kernel is found and from this
wave dispersion dependencies are obtained in 
explicit form.
 An equation of the  Wiener-Hopf type is also derived  and solved along the crack face, in order to compute the stress intensity factor for the semi-infinite crack. 
 The crack stability is analysed via the evaluation of the energy release rate for different contrasts in stiffness of the bonds.
\end{abstract}

\section{Introduction}\label{IntroLattice}

Models for cracks propagating in lattices have been extensively studied in \cite{Slepyan}. As noted in \cite{MarderGross}, discrete models enable one to answer fundamental questions about the influence of the micro-structure on the crack motion.

  It is possible to consider types of non-uniformity within a lattice,  which can bring new effects in the  wave dispersion and filtering properties of the structure. 
 These non-uniformities may be caused, for example, by thermal pre-stress of a constrained lattice whose ligaments have different coefficients of thermal expansion. Linearisation near the pre-stressed state may lead to a model of a lattice with contrasting stiffnesses of bonds.  Such an elastic lattice, containing a moving crack, is analysed in the present paper.  
 
 Problems involving  inhomogeneous lattices containing semi-infinite cracks  have been recently analysed. Examples include 
    those containing particles of contrasting mass  \cite{Mishetal2007}, and  
   the dynamic extraction of a chain within the lattice  \cite{Mishetal2008}.

The propagation of the crack may be caused by feeding waves, generated by some remote source, which lead to the breaking of subsequent bonds within the lattice. Feeding waves bring energy to the crack-front bonds which cause their disintegration one by one and this produces dissipative waves which carry energy away from the front.
 This process has been investigated in \cite{Slepyansq}, where   Mode III  crack propagation within a  square-cell lattice  is considered, under the assumption of  uniform straight line crack growth,   and steady state solutions have been obtained. 
  The lattice is assumed to have bonds of identical stiffness which  connect particles having a common mass.

 The analysis of Mode I and II  crack propagation in a uniform discrete triangular lattice was studied in \cite{Slepyantri}.  Compared to the case of the square-cell lattice, the equations of motion 
 relate components of the vector field of displacements.
 It 
 has been shown that  both the solution to this problem and   the explicit form of wave dispersion relations can be obtained.

 From 
 these lattice models, it is  possible to determine the regions of crack speeds for which we have steady state crack motion, 
and the stability of these states can also be determined 
by computing the energy release rate for the crack.  This has been examined in \cite{FineMard}, \cite{MarderGross} and \cite{MarderLiu}. 
Another application for a  lattice problem
is found in \cite{MIshetal2}.  Here, the model describing a structured interface along a crack with a harmonic feeding wave localised at the faces, was used to predict the position of the crack front. Numerical simulations were presented 
in \cite{knife, MIshetal2} showing that, for a given range of frequencies of the feeding wave, it was possible to have uniform crack growth or, in the non-linear regime of non-steady propagation, to identify an average crack speed, which is consistent with the prediction of the linear model linked to the crack propagating steadily.
  
The method of solution of these 
problems involves  formulating the 
 discrete lattice problem in terms of the Fourier transforms of functions describing the  displacements \cite{Mishetal}. A Wiener-Hopf functional equation is then derived along the crack faces and factorised to obtain the solution. The kernel in the Wiener-Hopf equation has the interpretation of the Fourier transform of the derivatives of Green's kernel. Similar features also appear in continuum models of cracks which are solved using singular integral equation techniques. [add reference]

\begin{figure}[htbp]
\centering
\includegraphics[width=0.6\textwidth]{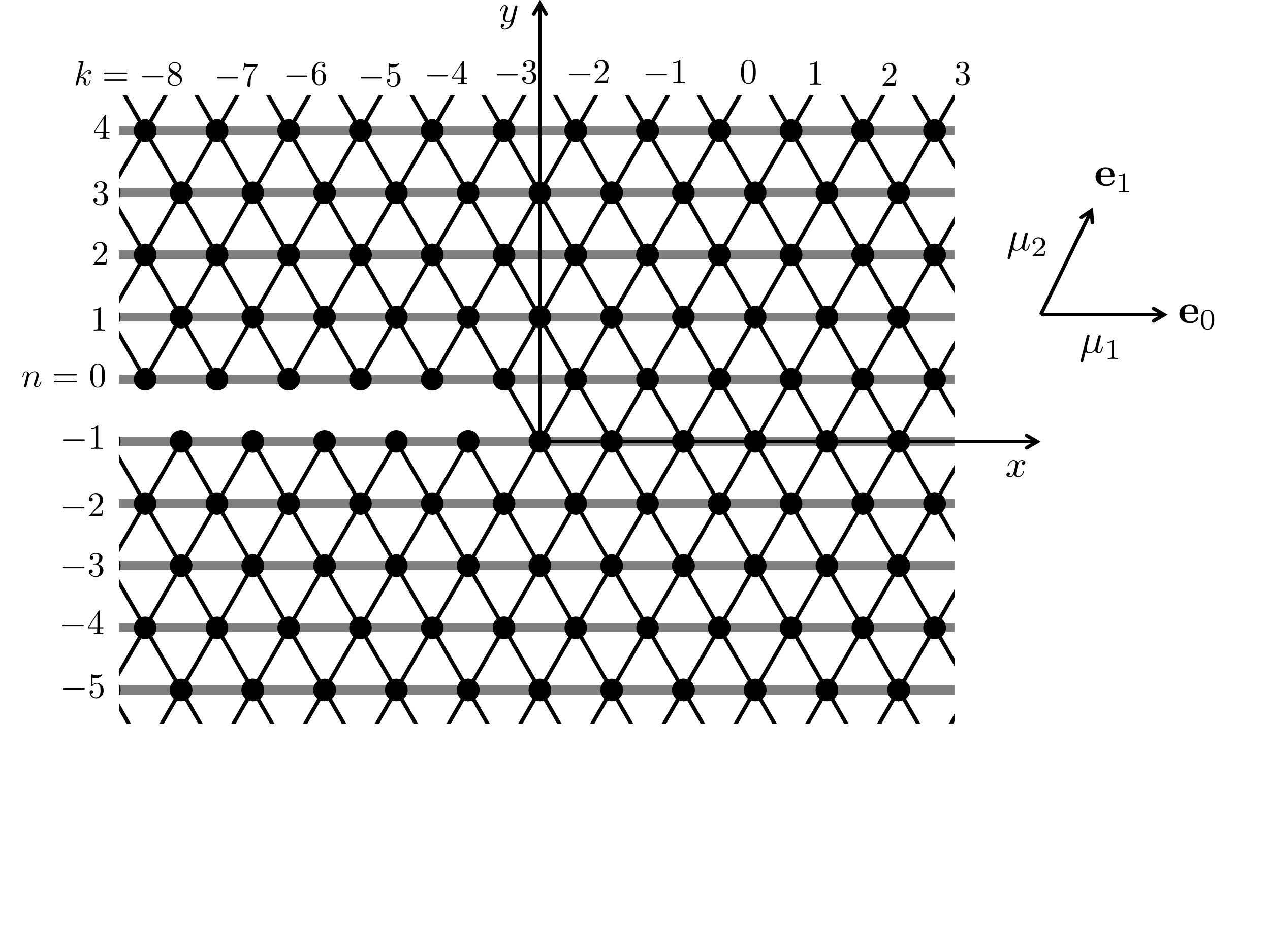}
\vspace{-0.7in}
\caption{The inhomogeneous lattice containing a semi-infinite crack showing the lattice coordinates. The crack is located between the rows $n=-1$ and $n=0$.}

\label{lattice_crack}
\end{figure}

In this article, using the method in \cite{Slepyantri}, our main goal is to solve the problem for a discrete triangular lattice, which is constrained at infinity,  containing a semi-infinite crack with additional  inhomogeneities being brought by the  inclined bonds having a contrast in stiffness to the horizontal bonds (see Figure \ref{lattice_crack}). 
 This contrast in stiffness may arise as an effect of thermal pre-stress, for instance, heating a lattice with bonds that have  different coefficients of  thermal expansion in the principal directions. The problem for a crack in the uniform triangular lattice has been studied in \cite[Chapter 12]{Slepyan}. 

The plan of the paper is as follows. section \ref{MainNotLattice} includes 
the description of the problem for a two dimensional lattice with a semi-infinite crack, containing particles of common mass with bonds having contrasting stiffnesses in the principal directions. In section \ref{solLatticeHF}, we rewrite the problem for the lattice using the continuous Fourier transform in the 
crack line direction, and these equations are solved for  the displacements of a particle inside the lattice. A  Wiener-Hopf equation for particles along the line of the crack is then written in section \ref{WHeqcrack} and the representation of  the kernel function of this  equation is derived.
An analysis of the dispersion relations obtained from the roots and poles of the kernel function is carried out in section \ref{RPLXI}. 
 In section \ref{SOLWH}, we solve the Wiener-Hopf equation of section \ref{WHeqcrack}. 
 In section \ref{ERR_sec}, we evaluate the energy release rate for the crack propagating through the inhomogeneous lattice
 and investigate its sensitivity to the crack speed and to the stiffness contrast.
We compute the stress intensity factor for the semi-infinite crack propagating through the inhomogeneous lattice and show its behaviour, for different stiffness contrasts, as a function of the  crack speed in section \ref{SIF_hom_approx}.  In section \ref{Conclusions1}, we give conclusions on the results presented here.

In addition to the main text of the article, we also provide Appendices which include more details of the derivations of the results contained in main body of the 
article. In Appendix I, the coefficient of the leading order term of the asymptotics for the kernel function near the zero wavenumber is studied.
Some comments relating to the analysis  of the poles for the kernel function of the Wiener-Hopf equation are presented in the Appendix II.  Properties of the kernel function which are necessary for its factorisation as a product of two  functions, analytic in different halves of the complex plane,  are proved in Appendix III. In Appendix IV, the derivation of the solution to the Wiener-Hopf equation is given. Finally, asymptotes in the vicinity of the zero wavenumber for the functions obtained after the factorisation of the kernel function are derived in Appendix V.


\section{Elastic lattice and the governing equations
}\label{MainNotLattice}

\subsection{The notion of Slepyan's crack}
The crack in the infinite triangular lattice considered here (see Figure \ref{lattice_crack}), is a semi-infinite fault which appears as the result of subsequent breakage of bonds in a straight line, and propagates with a constant speed $V$ according to $\eta=x-Vt$, (e.g. between rows $n=0$ and $-1$ in Figure \ref{lattice_crack}). Here, $x$ is the horizontal coordinate associated with the crack tip node, directed in the line of the propagation of the crack, and $t$ is time. This approach to describing the crack propagation, was first used in \cite{Slepyantri}.
 It allows for the reduction of the full transient problem to a simple difference problem in the vertical direction following the Fourier transform with respect to  $\eta$, where the forces in the  bonds of the lattice can be linked to the displacements and stresses through a Wiener-Hopf equation along the crack. Dispersion properties of this lattice configuration can also be obtained through the roots and poles of the kernel function involved this equation. 
The approach is elegant and leads to a closed form solution. One of the drawbacks is that the model does not serve the case of small crack velocities, as discussed further in the text below. However, for a finite range of values of a subsonic crack speed, one can identify a regime of steady crack propagation. In the real physical situation, crack propagation can be treated in the averaged sense, and the average crack velocity can be evaluated, as discussed in \cite{knife}.


\subsection{Non-uniform lattice}

We consider a triangular lattice containing a semi-infinite crack (see Figure \ref{lattice_crack}), where 
horizontal bonds are assumed to have stiffness $\mu_1$ and diagonal bonds  stiffness $\mu_2$. In the following, the ratio of bar stiffnesses is $\alpha=\mu_1/\mu_2$.
 The length of the bonds connecting the particles is normalised to  $1$. All particles are assumed to have identical mass $m$.

To characterise the positions of masses within the lattice, it is convenient to use 
the basis vectors 
$\mathbf{e}_0=(1, 0)^T$ and $\mathbf{e}_1=(1/2, \sqrt{3}/2)^T$.
The in-plane displacement vector 
 at the node $\Bx=k \mathbf{e}_0+n\mathbf{e}_1$, {where  $k, n$ are integers}, is   $\Bu(t, \Bx)$.
Projected displacements onto the horizontal and vertical axes, associated with the crack tip node, are denoted by $u$ and $v$, respectively.

We define the constant
\begin{equation}\label{subR}
C_R=\frac{1}{2m^{1/2}}\sqrt{ 2\mu_1+\mu_2-\sqrt{3\mu_1^2+(\mu_1-\mu_2)^2}}
\end{equation}
which will be determined, in section \ref{RPLXI}, as the Rayleigh wave speed for a  wave that propagates along the crack faces aligned with the $x$-axis,  which are traction free. The crack is assumed to propagate steadily through lattice with speed $V\le C_R$, and this propagation is caused by the breaking of the bonds between rows $n=-1$ and $n= 0$. 

\subsection{Normalised crack speed}
 Let $V_*$ be the dimensionless crack speed defined by $V_*=V\sqrt{m/\mu_2}$.
Since we consider sub-Rayleigh wave speed, using (\ref{subR}) we obtain 
\begin{equation}\label{eqg}
V_*^2 \le g(\alpha)= \frac{C_R^2 m}{\mu_2}=\frac{2\alpha+1-\sqrt{3\alpha^2+(\alpha-1)^2}}{4}\;.
\end{equation}
\begin{prop}\label{prog}
The function $g(\alpha)$ is increasing for $\alpha>0$, with $0< g(\alpha)\le 3/8$ and 
 $\displaystyle{\frac{dg}{d\alpha}(\alpha)< 3/4}$.
\end{prop}

\emph{Proof. }
For  $g(\alpha)$, in (\ref{eqg}), we have
\begin{equation*}\label{derg}
\frac{d g}{d \alpha}(\alpha)=\frac{1}{2}-\frac{4\alpha -1}{4\sqrt{3\alpha^2+(\alpha-1)^2}}\;,
\end{equation*}
and with  $0<\alpha\le 1/4$, it is clear that $g(\alpha)$ is an increasing function on this interval. Now, let $\alpha> 1/4$. Using the inequality 
\begin{equation}\label{eqest1}
3\alpha^2+(\alpha-1)^2 > \frac{(4\alpha-1)^2}{4}\;,
\end{equation}
leads to
\[\frac{d g}{d \alpha}(\alpha)
> 0, \quad \text{ for } \alpha>1/4
\;.\]
Therefore, $g(\alpha)$ is an increasing function for $\alpha >0$, and 
\[g(0)=0\;, \quad\text{ and }\quad g(\alpha)\le g(\infty)=\frac{3}{8}\;, \quad \alpha >0\;.\]
Next,
 owing to (\ref{eqest1})
\[\frac{d^2 g}{d\alpha^2}(\alpha)=\frac{1}{4\sqrt{3\alpha^2+(\alpha-1)^2}}\left(\frac{(1-4\alpha)^2}{3\alpha^2+(\alpha-1)^2}-4\right)< 0\;,\]
which shows $\displaystyle{\frac{dg}{d\alpha}(\alpha)}$ is a decreasing function, that takes its maximum as $\alpha\to 0$:
\[\frac{dg}{d\alpha}(0)=\frac{3}{4}\;.\]\hfill $\Box$

\subsection{Dynamic equations for the lattice above the crack}\label{deng0}
The equations of motion for a particle inside  the lattice ($n>0$) are:
\begin{eqnarray}\label{eqMotion}
m \ddot{\Bu}&=&
\mu_1\Be_0\cdot \Big\{\Bu\big|_{\Bx+\Be_0}
-\Bu\big|_{\Bx-\Be_0}-2\Bu\big|_\Bx\Big\}\Be_0
+\mu_2\sum_{j=1,2}\Bv_j \cdot \Big\{\Bu\big|_{\Bx+\Bv_j}+\Bu\big|_{\Bx-\Bv_j}-2\Bu\big|_{\Bx}\Big\}\Bv_j\;.
\end{eqnarray}
where $\Bv_1=\Be_1$ and $\Bv_2=\Be_1-\Be_0$.

Making a change of variable to the moving coordinate system, we   set $\eta=x-Vt$. Then  (\ref{eqMotion}) 
in terms of projected displacements becomes
\begin{eqnarray*}&&\frac{mV^2}{\mu_2}\frac{d^2 u}{d\eta^2}(\eta, n)
-
\frac{1}{4}[u(\eta+1/2, n+1)+u(\eta-1/2, n+1)+u(\eta+1/2, n-1)+u(\eta-1/2, n-1)]\\ \\
&&-\frac{\sqrt{3}}{4}[v(\eta+1/2, n+1)-v(\eta-1/2, n+1)
+v(\eta-1/2, n-1)-v(\eta+1/2, n-1)]\\ \\&&-\alpha [u(\eta+1, n)+u(\eta-1, n)]
+(1+2\alpha) u(\eta,n)=0\;,
\end{eqnarray*}
and
\begin{eqnarray*}
&&\frac{mV^2}{\mu_2}\frac{d^2 v}{d\eta^2}(\eta, n)-\frac{\sqrt{3}}{4}[u(\eta+1/2, n+1)-u(\eta-1/2, n+1)
-u(\eta+1/2, n-1)+u(\eta-1/2, n-1)]\\ \\
&&-\frac{3}{4}[v(\eta+1/2, n+1)+v(\eta-1/2, n+1)
+v(\eta+1/2, n-1)+v(\eta-1/2, n-1)]+3v(\eta, n)=0\;,
\end{eqnarray*}
where $u(\eta,n)$ and $v(\eta, n)$ are the components of the displacement vector $\Bu$, along row $n$  after the change of variable.

\section{Solution for the lattice half-plane}\label{solLatticeHF}
Here, for the intact part of the  lattice $(n>0)$ we will obtain the solution to the problem of section \ref{deng0}. We follow  the method discussed in \cite{Slepyantri}, which  makes use of the continuous Fourier transforms $u^F$ and $v^F$ with respect to  the moving coordinate  associated with the crack.
\subsection{The continuous Fourier transform of the system}
We take the Fourier transform with respect to $\eta$, and obtain
\begin{eqnarray}
[u^F(\xi, n+1)+u^F(\xi,n-1)] \cos(\xi/2)
-2[1+2\alpha(1-\cos(\xi))+m(\varepsilon+{\rm i} \xi V)^2/\mu_2 ]u^F(\xi,n)
\nonumber&& 
\\ \qquad\qquad\qquad
-\text{i} \sqrt{3}\sin(k/2)[v^F(\xi, n+1)-v^F(\xi, n-1)]&=&0\;,\label{FeqF1}
\end{eqnarray}
\begin{eqnarray}
3[v^F(\xi, n+1)+v^F(\xi,n-1)]\cos(\xi/2)
 -2[3+m(\varepsilon+{\rm i} \xi V)^2/\mu_2] v^F(\xi, n)
 \nonumber
&& \\ \qquad\qquad
-\text{i} \sqrt{3} \sin(\xi/2)[u^F(\xi, n+1)-u^F(\xi, n-1)]&=&0\;,\label{FeqF2}
\end{eqnarray}
where $\varepsilon$ is the regularization parameter, $0<\varepsilon\ll 1$.
\subsection{General solution for the half-plane lattice}
The functions $u^F$ and $v^F$ are sought in the form
\[u^F=C \Lambda^n\quad \text{and}\quad v^F=D \Lambda^n\;.\]
Insertion of these into (\ref{FeqF1}) and (\ref{FeqF2}) yield
\begin{eqnarray*}
C\{(\Lambda+1/\Lambda) \cos(\xi/2)-2[1
+2
\alpha
(1-\cos(\xi))+m(\varepsilon+{\rm i} \xi V)^2/\mu_2]\}
-\text{i}
\sqrt{3}\sin(\xi/2)(\Lambda-1/\Lambda)D &=&0\;,\label{FeqC1}
\end{eqnarray*}
\begin{eqnarray*}
D\{3
(\Lambda+1/\Lambda)\cos(\xi/2)-2(3
+Y)\}
-\text{i} 
\sqrt{3} \sin(\xi/2)(\Lambda-1/\Lambda)C&=&0\;.\label{FeqC2}
\end{eqnarray*}
For non-trivial solutions of $C$ and $D$, the biquadratic equation in terms of $\Lambda$ must be satisfied:
\begin{eqnarray}
(\Lambda+1/\Lambda)^2-2 
 \{2
 +2
 \alpha (1-\cos(\xi))+\frac{4}{3\mu_2}m(\varepsilon+{\rm i} \xi V)^2\}(\Lambda+1/\Lambda)  \cos(\xi/2)&&\nonumber\\+ 2
 \{
  1+\cos(\xi)+4\alpha(1-\cos(\xi))\}+\frac{4}{3\mu_2}m(\varepsilon+{\rm i} \xi V)^2 \{4
  +2\alpha
  (1-\cos(\xi))+\frac{m}{\mu_2}(\varepsilon+{\rm i} \xi V)^2\}&=&0\;.\label{bilambda}
\end{eqnarray} 
We require that $|\Lambda|\le 1$, and this leads to the solutions
\begin{eqnarray}
&&\Lambda_j
=z_j
\pm\sqrt{z_j
^2-1}\;, \quad j=1,2,\label{biquadroots}
\end{eqnarray}
where the sign in front of the square root in $\Lambda_j$
has to be chosen so that the condition $|\Lambda_j
|\le 1$, $j=1,2$, for $\xi \in \mathbb{R}$ and $V > 0$, is satisfied. 
 For 
  $j=1,2$
\begin{eqnarray} &&z_{j}=\Big(1+2\alpha\sin^2(\xi/2)+\frac{2m(\varepsilon+{\rm i} \xi V)^2}{3\mu_2}\Big)\cos(\xi/2)\nonumber
\nonumber \\&& +(-1)^j \Bigg[{\frac{m^2(\varepsilon+{\rm i} \xi V)^4}{9\mu_2^2}-4\sin^2(\xi/2) \left(
\alpha \sin^2(\xi/2)+\frac{m(\varepsilon+{\rm i} \xi V)^2}{3\mu_2}\right)\nonumber
}\\&&\times {\left(
\alpha \sin^2(\xi/2)+\frac{m(\varepsilon+{\rm i} \xi V)^2}{3\mu_2}+1
-
\alpha \right)}\Bigg]^{1/2}
\;. 
\label{eqzj}
\end{eqnarray}
Then for $n>0$, with such choices of $\Lambda$, the Fourier transforms of the displacements have the form
\begin{equation}\label{eqUyF}
\left(\begin{array}{c}
u^F(\xi,n)\\ \\
v^F(\xi, n)
\end{array}\right)
=\frac{1}{f_v(\Lambda_1)f_v(\Lambda_2)}\left(\begin{array}{cccc}f_v(\Lambda_1)f_v(\Lambda_2)& && f_v(\Lambda_1)f_v(\Lambda_2)\\ \\
-{\rm i}f_u(\Lambda_1)f_v(\Lambda_2)&& &-{\rm i}f_u(\Lambda_2)f_v(\Lambda_1)\end{array}\right)\left(\begin{array}{c}C_1\Lambda_1^n\\ \\ C_2\Lambda_2^n\end{array}\right)
\end{equation}
where
\[f_u(\Lambda)=-\sqrt{3}
 \sin(\xi/2) (\Lambda^2-1) \quad \text{and} \quad 
 f_v(\Lambda)=3
 (\Lambda^2+1)\cos(\xi/2)-2\Lambda(3
 +m(\varepsilon+{\rm i} \xi V)^2/\mu_2)\;. \]

\subsection{Mode I Symmetry conditions}
For Mode I, using the symmetry relations
\begin{equation}\label{symmetrycond}
\begin{array}{c}
u(\eta, -n-1)=u(\eta, n)\;,\qquad v(\eta, -n-1)=-v(\eta, n)\;,\end{array}
\end{equation}
 we can extend the solution in the upper half-plane
  (\ref{eqUyF})
   to the lower half-plane. 
Then, for $n \le -1$, the Fourier transforms of the horizontal and vertical displacements are
\begin{equation*}
\left(\begin{array}{c}
u^F(\xi,n)\\ \\
v^F(\xi, n)
\end{array}\right)
=\frac{1}{f_v(\Lambda_1)f_v(\Lambda_2)}\left(\begin{array}{cccc}f_v(\Lambda_1)f_v(\Lambda_2)& && f_v(\Lambda_1)f_v(\Lambda_2)\\ \\
{\rm i}f_u(\Lambda_1)f_v(\Lambda_2)&& &{\rm i}f_u(\Lambda_2)f_v(\Lambda_1)\end{array}\right)\left(\begin{array}{c}C_1\Lambda_1^{-n-1}\\ \\ C_2\Lambda_2^{-n-1}\end{array}\right)\;.
\end{equation*}
 The forces in the bonds directed along $-\mathbf{e}_1$ and $\mathbf{e}_0-\Be_1$, respectively, for a node in the upper half-plane of the lattice, have the form

\begin{equation}\mathcal{S}(\eta, n)=\frac{1}{2}[u(\eta, n)-u(\eta-1/2, n-1)]
+\frac{\sqrt{3}}{2}[v(\eta, n)-v(\eta-1/2, n-1)]\;,
\label{Q4}
\end{equation}
\begin{equation}
\mathcal{T}(\eta, n)
=
\frac{1}{2}[u(\eta+1/2, n-1)-u(\eta, n)]
+\frac{\sqrt{3}}{2}[v(\eta, n)-v(\eta+1/2, n-1)]\;.
\label{Q5}
 \end{equation}
The conditions (\ref{symmetrycond}) also give 
\[\mathcal{S}(\eta, 0)=\frac{1}{2}[u(\eta, 0)-u(\eta-1/2,0)]+\frac{\sqrt{3}}{2}[v(\eta, 0)+v(\eta-1/2, 0)]\;, \quad\mathcal{T}(\eta, 0)=\mathcal{S}(\eta+1/2, 0)\]
or
\begin{equation}\label{Q4Frep}
\mathcal{S}^F(\xi, 0)=\frac{1}{2}(1-e^{\text{i}\xi/2})u^F(\xi, 0)+\frac{\sqrt{3}}{2}(1+e^{\text{i}\xi/2})v^F(\xi, 0)\;,
\quad \mathcal{T}^F(\xi, 0)=\mathcal{S}^F(\xi, 0)e^{-\text{i}\xi/2}\;.
\end{equation}
The representation
 (\ref{eqUyF}), 
  then allows $\mathcal{S}^F$ in (\ref{Q4Frep}) to be rewritten as
\begin{equation}\label{eqQ4Frep2}
\mathcal{S}^F(\xi, 0)=\frac{1}{2f_v(\Lambda_1) f_v(\Lambda_2)} \BR^T \BT\BB(\Lambda_1, \Lambda_2)\BC\;, 
\end{equation}
with
\[\BC=\left(\begin{array}{c} C_1\\ \\ C_2\end{array}\right)\;, \qquad \BR=\left(\begin{array}{c}
1-e^{\text{i}\xi/2}\\ \\ \text{i}(1+e^{\text{i} \xi/2})\end{array}\right)\;,\]
and
\[\BT=\left(\begin{array}{cc}1 & 0\\
0 &\sqrt{3}
\end{array}\right)\;,\qquad \BB(\Lambda_1, \Lambda_2)=\left(\begin{array}{cccc}
f_v(\Lambda_1)f_v(\Lambda_2) &&&f_v(\Lambda_1)f_v(\Lambda_2)\\
\\
-f_u(\Lambda_1) f_v(\Lambda_2)& && -f_u(\Lambda_2)f_v(\Lambda_1) 
\end{array}\right)\;.\]
\section{The Wiener-Hopf equation along the crack}\label{WHeqcrack}

In this section, we will use the equations of motion for a particle on the upper crack face 
($n=0$), 
 to derive a Wiener-Hopf  equation. The kernel function of this equation is also obtained and its properties are investigated. This function is used in $(a)$ section \ref{RPLXI} for analysing the dispersion properties of the lattice 
 and $(b)$ for computing the energy release rate of the propagating crack in section \ref{ERR_sec}.
\subsection{Dynamic equations along the crack face $(n=0)$}
\begin{figure}[htbp]
\centering
\includegraphics[width=0.45\textwidth]{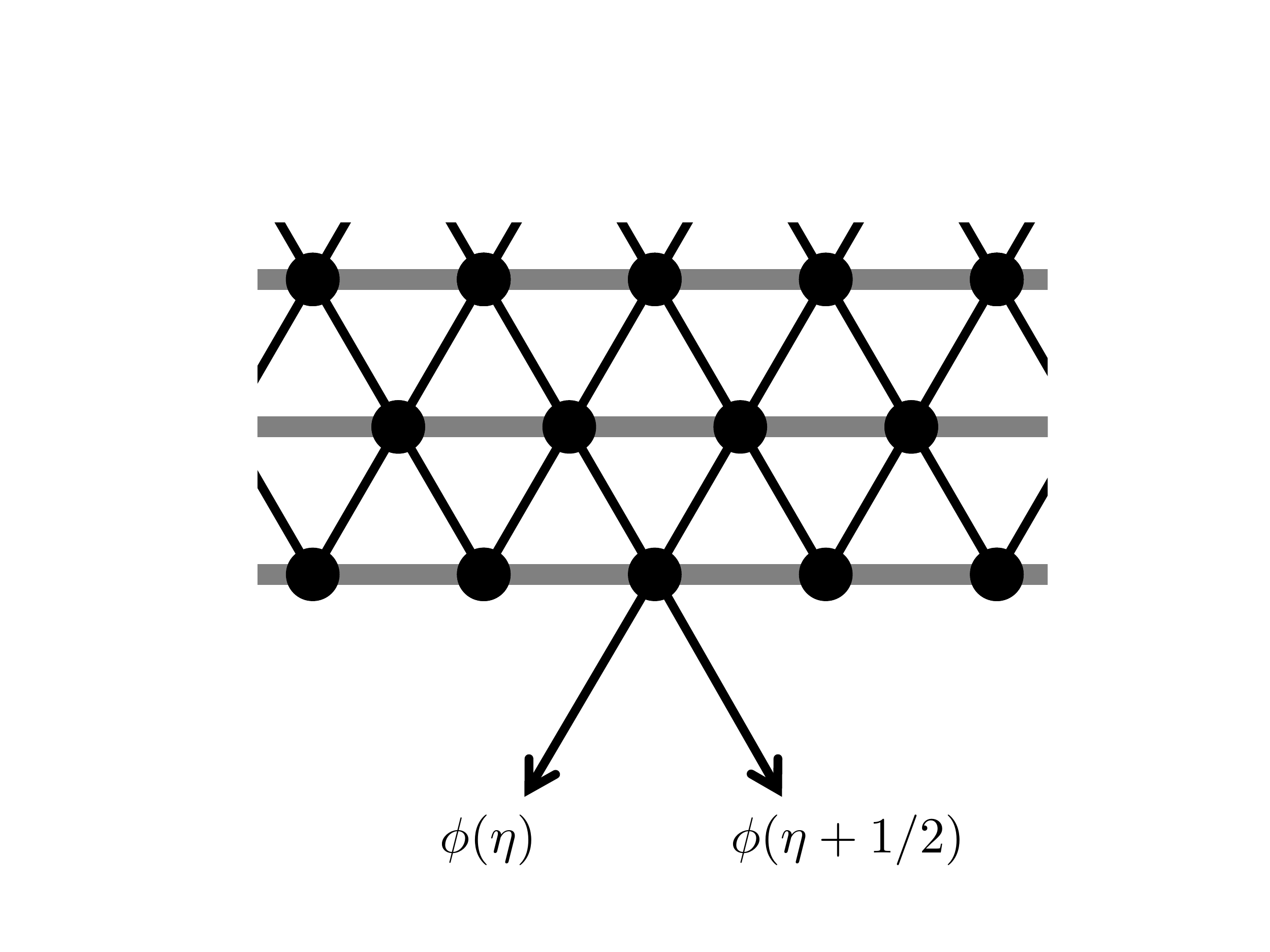}

\caption{
The particles along the crack face at $n=0$, acted on by external forces $\phi(\eta)$ and $\phi(\eta+1/2)$.}
\label{elem_cell_pic}
\end{figure}

Now we consider the equations of motion for particles along $n=0$. The equations are similar to those presented in  section \ref{deng0}. However, since particles along the crack face have no bonds corresponding  to the  vectors $-\Be_1$ and $\mathbf{e}_0-\Be_1$, and are acted on by an external force $\phi$ (see Figure \ref{elem_cell_pic}), this should be taken into account. 

Ahead of the crack, for particles along $n=0$, the elongation of the bonds 
corresponding to $\mathbf{e}_0$, $\Be_1$, $\Be_1-\Be_0$ and $-\Be_0$ are the same as those for a particle in the intact upper lattice half plane. 
 Also 
  the internal force of the bond in the direction of $-\mathbf{e}_1$, can be written as
 \begin{equation}\label{P4F}
 \mathcal{M}(\eta, 0)=\mathcal{S}(\eta)H(\eta), \qquad 
 \mathcal{M}^F=\mathcal{S}_+\;,\end{equation}
 where $H$ is the Heaviside function.  Using (\ref{Q4Frep}) and $
 \mathcal{M}^F$, a similar expression for the internal force in the bond directed along $\mathbf{e}_0-\Be_1$ is obtained, $
 \mathcal{N}^F=\mathcal{S}_+e^{-\text{i}k/2}$. The external forces acting on the particles on the line $n=0$, from (\ref{Q4Frep}), have the form $\phi(\eta)$ and $\phi(\eta+1/2)$, which are directed along $-\mathbf{e}_1$ and $\mathbf{e}_0-\Be_1$, respectively.  

Then, the Fourier transform of  the equations of motion for the particles along the line $n=0$ are
\begin{eqnarray*}
\frac{m}{\mu_2}(\varepsilon+{\rm i}\xi V)^2 u^F(\xi, 0)&=&
\frac{1}{2}\cos(\xi/2)u^F(\xi, 1)
-\{2\alpha(1-\cos(\xi/2))+1/2\}u^F(\xi, 0)\\
&&
-\frac{{\rm i} \sqrt{3}}{2} \sin(\xi/2) v^F(\xi, 1)
-\frac{1}{2}(1-e^{-\text{i} k/2})(\mathcal{S}_++\phi^F)\;,
\end{eqnarray*}
\begin{eqnarray*}
\frac{m}{\mu_2}(\varepsilon+{\rm i}\xi V)^2  v^F(k, 0)&=&
-\frac{{\rm i}\sqrt{3}}{4} \sin(\xi/2)u^F(\xi, 1)+\frac{3}{4}\cos(\xi/2) v^F(\xi, 1)
-\frac{3}{2}v^F(\xi, 0)-\frac{\sqrt{3}}{2}(1+e^{-\text{i}k/2}) (\mathcal{S}_++\phi^F)\;.
\end{eqnarray*}
In order that the expressions
  (\ref{eqUyF}) fulfill
 these  equations, the system 
 \begin{equation*}\label{con_sys}
\begin{array}{c} \displaystyle{\mathcal{T}^F(\xi, 0)-\mathcal{S}^F(\xi, 0)=-(1-e^{-\text{i} \xi/2})(\mathcal{S}_++\phi^F)\;,}\qquad
 \displaystyle{\mathcal{T}^F(\xi, 0)+\mathcal{S}^F(\xi, 0)=(1+e^{-\text{i} \xi/2})(\mathcal{S}_++\phi^F)\;,}
 \end{array}
 \end{equation*}
 should be satisfied (see  (\ref{FeqF1}), (\ref{FeqF2}) and (\ref{Q4Frep})).
  Using the representation  
 (\ref{eqUyF})
for the displacements when $n\ge 0$, 
in the above, leads to
\[\BGX(\Lambda_1, \Lambda_2) \BC=-\Lambda_1 \Lambda_2 f_v(\Lambda_1) f_v(\Lambda_2)\overline{\BR}(\mathcal{S}_++\phi^F)\;.\]
where the matrix $\BGX(\Lambda_1, \Lambda_2)=[\GX_{ij}(\Lambda_1, \Lambda_2)]_{i, j=1}^2$ has the entries
\[\GX_{11}(\Lambda_1, \Lambda_2)=\Lambda_1 f_v(\Lambda_2)(f_v(\Lambda_1)(\cos(\xi/2)-\Lambda_1)+\sqrt{3} \sin(\xi/2) f_v(\Lambda_1))\;,\]
\[\GX_{21}(\Lambda_1, \Lambda_2)=\Lambda_2 f_v(\Lambda_1)(-\sqrt{3}f_u(\Lambda_1)(\cos(\xi/2)-\Lambda_1)+f_v(\Lambda_1) \sin(\xi/2)) \;,\]
\[\GX_{12}(\Lambda_1, \Lambda_2)=\GX_{11}(\Lambda_2, \Lambda_1) \quad \text{and}\quad \GX_{22}(\Lambda_1, \Lambda_2)=\GX_{21}(\Lambda_2, \Lambda_1)\;.\]
Therefore
\[ \BC=-\Lambda_1 \Lambda_2 f_v(\Lambda_1) f_v(\Lambda_2)[\BGX(\Lambda_1, \Lambda_2)]^{-1}\overline{\BR}(\mathcal{S}_++\phi^F)\;.\]
Combining this with (\ref{eqQ4Frep2}) yields
\begin{equation}\label{FEQQ41}
\mathcal{S}^F(\xi, 0)=-\frac{\Lambda_1 \Lambda_2}{2\text{det}(\BGX(\Lambda_1, \Lambda_2))} \BR^T\BGO(\Lambda_1, \Lambda_2)\overline{\BR}(\mathcal{S}_++\phi^F)\;.\end{equation}
with
\[\BGO(\Lambda_1, \Lambda_2)= \BT\BB(\Lambda_1, \Lambda_2)\text{adj}(\BGX(\Lambda_1, \Lambda_2))\;,\]
and 
\begin{equation}\label{detXI}
\begin{array}{c}
\displaystyle{\text{adj}(\BGX(\Lambda_1, \Lambda_2))=\left(\begin{array}{ccc}
\GX_{21}(\Lambda_2, \Lambda_1) & &-\GX_{11}(\Lambda_2, \Lambda_1)\\\\
-\GX_{21}(\Lambda_1, \Lambda_2) && \GX_{11}(\Lambda_1, \Lambda_2) 
\end{array}\right)\;, }\\  \\
\displaystyle{\text{det}(\BGX(\Lambda_1, \Lambda_2))=\GX_{11}(\Lambda_1, \Lambda_2)\GX_{21}(\Lambda_2, \Lambda_1)-\GX_{11}(\Lambda_2, \Lambda_1)\GX_{21}(\Lambda_1, \Lambda_2)\;.}
\end{array}
\end{equation}
Since 
\[\text{adj}(\BGX(\Lambda_2, \Lambda_1))=-\left(\begin{array}{cc}0&1 \\1&0\end{array}\right)\text{adj}(\BGX(\Lambda_1, \Lambda_2))\;,\quad\BB(\Lambda_2, \Lambda_1)=\BB(\Lambda_1, \Lambda_2)\left(\begin{array}{cc}0&1\\1&0
\end{array}\right)\;,\]
the matrix $\BGO$ matrix satisfies the relation
\begin{equation}\label{OMantsym}
\BGO(\Lambda_1, \Lambda_2)=-\BGO(\Lambda_2, \Lambda_1)\;.
\end{equation}
\subsection{The Wiener-Hopf equation}
Next we set 
\[\mathcal{S}^F(\xi, 0)=\mathcal{S}_++\mathcal{S}_-\;,\]
where $\mathcal{S}_+$, $\mathcal{S}_-$ are functions analytic in the upper and lower half-planes of the complex plane $\xi$, and as in (\ref{P4F}), $\mathcal{M}^F(\xi, 0)=\mathcal{S}_+$, whereas the quantity $\mathcal{S}^F$ satisfies (\ref{eqQ4Frep2}).
After inserting this into (\ref{FEQQ41}), we obtain an equation of the form
\begin{equation}\label{eqWH}
L(\xi) \mathcal{S}_++\mathcal{S}_-=(1-L(\xi))
\phi^F\;.
\end{equation}
Here, 
\[
L(\xi)=1+
\frac{\Lambda_1 \Lambda_2}{2\text{det}(\BGX(\Lambda_1, \Lambda_2))} \BR^T\BGO(\Lambda_1, \Lambda_2)\overline{\BR}\;,\]
where $\Lambda_j$, $j=1,2,$ are given in (\ref{biquadroots}).

We note that due to (\ref{OMantsym}) and the antisymmetry of $\text{det}(\BGX(\Lambda_1, \Lambda_2))$ in (\ref{detXI}) with respect to $\Lambda_1$ and $\Lambda_2$, that $L(\xi)$ is symmetric with respect to $\Lambda_1$ and $\Lambda_2$. 
We put 
\[\Lambda_j^*=z_j-\sqrt{z_j^2-1}\;, j=1,2\;,\]
and then
an equivalent representation for $L(\xi)$ in terms of $z_1$ and $z_2$ (see  (\ref{biquadroots}) and (\ref{eqzj})) is given by
\begin{equation}\label{Lkn1n2}
L(\xi)=\frac{3r(\Lambda_1^*)r(\Lambda_2^*)(z_2-z_1)\sqrt{z_1^2-1}\sqrt{z_2^2-1}}{r(\Lambda_1^*)F(z_2)\sqrt{z_1^2-1}-r(\Lambda_2^*)F(z_1)\sqrt{z_2^2-1}}\;,
\end{equation}
where
\[r(\Lambda)=\left\{\begin{array}{ll}
1,& \quad|\Lambda|\le 1\;,\\ \\
-1,&\quad \text{otherwise,}\end{array}\right.\]
\begin{eqnarray*}
F(z)&=&3(\cos(\xi/2)-z)^2+6\alpha \sin^2(\xi/2)(1+\cos(\xi/2))(1-z)\\
&&+\frac{m}{\mu_2}(\varepsilon+{\rm i}\xi V)^2[1-z\cos(\xi/2) +(1+\cos(\xi/2))(1-z)]\;.\end{eqnarray*}
The representation for $L(\xi)$ in (\ref{Lkn1n2}), contains the stiffness contrast parameter $\alpha$ which is found in the function $F$. For the case $\alpha=1$,  (\ref{Lkn1n2}) is  similar to (12.47) of \cite[Section 12.3.5]{Slepyan}, with the only difference here  being that the functions $r(\Lambda_1^*)$, $r(\Lambda_2^*)$ have been included to indicate the change of the branch cuts of the square roots whenever $|\Lambda_{1,2}|>1$.
 
In the derivation of (\ref{Lkn1n2}), from (\ref{biquadroots}) we used 
\[\Lambda_j=z_j-r(\Lambda_j^*)\sqrt{z_j^2-1}\;,\]
together with fact
\[\frac{1}{\Lambda_j}=z_j+r(\Lambda_j^*)\sqrt{z_j^2-1}\;,  \quad \text{ for }j=1,2\;.\]

\subsection{Asymptotics of $\Lambda_{1,2}$ }
Here, we obtain estimates for the behaviour of 
the roots of (\ref{bilambda}), in the neighbourhood of zero and infinity, and discuss their properties.

For the auxiliary functions $z_j$, $j=1, 2$ we have
\begin{equation}\label{infasyn12}
z_{j}=\frac{m(\varepsilon+{\rm i}\xi V)^2}{3\mu_2}\left[2\cos({\xi}/{2})+(-1)^j\sqrt{1-4\sin^2({\xi}/{2})}\right]+O(Y^{1/2}_q),  \quad j=1,2\;, \quad \xi\to\pm\infty\;,
\end{equation}
where the term in the parenthesis is
non-zero for all $\xi$.  Here, $Y_q=O((\varepsilon+{\rm i}\xi V)^2)$ for most values of $\xi$,
 however for a periodic sequence of $\xi=\xi_m$ we have this is $Y_q=O(1)$.

In  the other limit, $\xi \to 0$, we obtain 
\[z_{j}=1+
\frac{1}{24}d_j\xi^2+O(\xi^4)\;,\quad \xi \to 0\;,\]
\begin{equation}\label{eqsqasy}\sqrt{z_j^2-1}=\frac{\sqrt{3}\xi}{6} \sqrt{d_j}+O(|\xi|^3), \quad j=1, 2\;,
\end{equation}
where
\begin{equation}\label{rad1}
d_j=3(4\alpha-1)
-{16V_*^2} +(-1)^j4M_0\;,\quad j=1, 2\;,
\end{equation}
\begin{equation}\label{M0}
M_0=\sqrt{{4 V^4_*}+3(\alpha-1) \left(3\alpha -{4V^2_*}\right)}\;,
\end{equation}
and $V_*=V\sqrt{m/\mu_2}$, as above.

Now we consider the behaviour of $\Lambda_{1, 2}$  at infinity,
we have 
\begin{equation}\label{eqlam1a}
\Lambda_j=\frac{1}{2z_j}+O(|z_j|^{-3})\;, \quad |\xi|\to \infty\;,
\end{equation}
that is, $\Lambda_j=O(\xi^{-2})$ as $|\xi|\to \infty$ and  
here the sign in front of the square root in (\ref{biquadroots}) is chosen to maintain the condition $|\Lambda_j| \le 1$, $j=1,2$, in this limit. 

Near zero
\begin{equation}\label{eqlam1b}
\Lambda_{j}=1\pm 
\frac{\sqrt{3}\xi}{6} 
\sqrt{d_j}
+O(\xi^2)\;, \quad \xi \to 0\;. 
\end{equation}
In order that the correct sign in front of the above square root is chosen  (so that $|\Lambda_{1,2}|\le 1$), it is necessary to 
analyse the behaviour of $d_j$, $j=1,2$ in (\ref{rad1})
for $\alpha>0$ and $0<V_*\le C_R(m/\mu_2)^{1/2}$.

\subsection*{The constants $d_{1,2}$}

Here we show that $d_{1, 2}$ do not belong to the negative real axis in the complex plane. Otherwise
\[1+\frac{\xi^2}{12}|d_j|>1\;,\]
and then higher order terms would be needed in (\ref{eqlam1b}) to determine if $|\Lambda_{1,2}|\le 1$ as $\xi\to 0$.
 The proposition below is proved for $d_1$ and the result extends to $d_2$.

\begin{prop}\label{propd1}
For $\alpha>0$ and $0<V_*\le C_R(m/\mu_2)^{1/2}$, we have $d_1 \in \mathbb{C}\backslash \mathbb{R}_-$.
\end{prop}

\emph{Proof. }We begin by  assuming on the contrary that  $d_1$ is negative, which is equivalent to the inequality 
\begin{equation}\label{validineq1}3(4\alpha-1)
-{16V_*^2} <4\sqrt{{4 V^4_*}+3(\alpha-1) \left(3\alpha -{4V^2_*}\right)}
\end{equation}
with the assumption that the term under the square root in the right-hand side is positive. This gives rise to the cases: 
\begin{enumerate}[a)]

\item 
\[3(4\alpha-1)
-{16V_*^2}>0\;,\]
together with
\[(3(4\alpha-1)
-{16V_*^2})^2 <16({4 V^4_*}+3(\alpha-1) \left(3\alpha -{4V^2_*}\right))\]
or

\item \[3(4\alpha-1)
-{16V_*^2}<0 \quad \text{ and }\quad {4 V^4_*}+3(\alpha-1) \left(3\alpha -{4V^2_*}\right)>0\;.\]

\end{enumerate}



\emph{The Case a$)$. }One can obtain the inequality
\[V_*^4-\frac{2\alpha+1}{2}V_*^2+\frac{3}{8}\left(\alpha+\frac{1}{8}\right)<0\;.\]
The quadratic function of $V_*^2$, in the left-hand side, has the zeros 
\[v_{1,2}^2=\frac{3}{8}\;, \frac{1}{8}+\alpha\;,\]
which for $\alpha>0$, are both positive. Then (\ref{validineq1}) is only valid if $V_*^2 \in (v_1^2, v_2^2)$ for $\alpha \ge 1/4$ or $V_*^2 \in (v_2^2, v_1^2)$ for $\alpha < 1/4$. We now show $V_*^2$ lies outside these intervals for $\alpha>0$. 

By Proposition \ref{prog}, 
\begin{equation}\label{maxeq1}
\max\{V_*^2\} =g(\alpha)\le v_1^2\;, \quad\text{ when }\quad \alpha\ge1/4\;,
\end{equation}
where $g(\alpha)$ is given in (\ref{eqg}).

For $\alpha <1/4$, owing again to Proposition \ref{prog}, 
\[\frac{dg}{d\alpha} (\alpha)<\frac{3}{4}\;, \quad \text{ for } \quad \alpha>0 \;, \quad  g(0)< \frac{1}{8}, \quad \text{ and }\quad g(1/4)=\frac{3}{8}\Big(1-\frac{\sqrt{3}}{3}\Big)<\frac{3}{8}\;. \]
Thus 
\[\max\{V_*^2\} =g(\alpha)
< v_2^2\;, \quad\text{ when }\quad 0< \alpha<1/4\;.\]
This inequality with (\ref{maxeq1}) shows that (\ref{validineq1}) is not valid for $V_*\in (0, C_R\sqrt{m/\mu_2}]$ and $\alpha>0$ for case $a)$ above.


\vspace{0.1in}\emph{The Case b$)$. }The second inequality of this case can be rewritten as
\begin{equation}\label{valideq2}
V_*^4+3(1-\alpha)V_*^2-\frac{9\alpha(1-\alpha)}{4}>0\;,
\end{equation}
where now the quadratic function on the left-hand side has zeros
\[w_{\pm}^2=\frac{3}{2}[\alpha-1 \pm \sqrt{1-\alpha}]\;.\]

\vspace{0.1in}When $\alpha\ge 1$ 
the inequality (\ref{valideq2}) is satisfied (since the zeros of the left-hand side are complex), and it remains to show   that the inequalities 
\begin{equation}\label{con1}
V_*^2 \le 
\frac{C_R^2m}{\mu_2}\quad \text{ and } \quad V_*^2 >\frac{3(4\alpha-1)}{16}\;, 
\end{equation}
do not hold.  These 
 lead to
\[\frac{3(4\alpha-1)}{16}< g(\alpha)\;,\quad
\text{  or } \quad
\sqrt{3\alpha^2+(\alpha-1)^2} < \frac{7}{4}-\alpha\;,\quad\]
where the definition of $g(\alpha)$ in  (\ref{eqg}) has been used. From the above inequalities, we have a contradiction to (\ref{con1}) for $\alpha \ge 7/4$.
Assume $1\le \alpha<7/4$, then the last inequality leads to 
\[\alpha^2+\frac{1}{2}{\alpha}-\frac{11}{16}<0\;.\]
This is valid for $\alpha \in (0,(-1/2+\sqrt{3})/2) $. Hence (\ref{validineq1}) does not hold.

 \vspace{0.1in}When 
 {$\alpha<1$,} for  (\ref{valideq2}) to be satisfied we put 
 \[V_*^2>w_+^2=\frac{3}{2}[\alpha-1 + \sqrt{1-\alpha}]\]
 where
 \[V_*^2 \le \frac{C_R^2m}{\mu_2}
 \;,\quad \text{ and }\quad V_*^2 > \frac{3(4\alpha-1)}{16}\;. \]
 Here, as before, the last two inequalities are valid if $0<\alpha<(-1/2+\sqrt{3})/2$. It remains to check that
 \begin{equation}\label{eqalphin}V_*^2>w_+^2=\frac{3}{2}[\alpha-1 + \sqrt{1-\alpha}]\quad \text{ and } 
 \quad V_*^2 \le \frac{C_R^2m}{\mu_2}\;.
 \end{equation}
Both of these yield  the inequality
 \[\alpha^2 (\alpha^2+\alpha-\frac{7}{4}) > 0\;.\]
 This
  is valid for $\alpha>-1/2+\sqrt{2}$
and, since we have  assumed that $0<\alpha<(-1/2+\sqrt{3})/2$, we have a contradiction to (\ref{eqalphin}).
 Therefore, (\ref{validineq1}) is not valid for $V_*\in (0, C_R(m/\mu_2)^{1/2}]$ and $\alpha>0$ in case $b)$ above. \hfill $\Box$

\subsubsection{Asymptotes of $L(\xi)$}
Here, we obtain the asymptotics of $L(\xi)$  for $\xi\to 0$. 

\begin{prop}\label{Propasy0}
For $\xi\to 0$, we have
\begin{equation}\label{eqLasy}
L(\xi)\sim  L_0\frac{1}{\sqrt{(\varepsilon+{\rm i} \xi)(\varepsilon-{\rm i} \xi)}}+O(|\xi|)\;,\qquad \varepsilon \to +0\;,
\end{equation}
where
\begin{equation}\label{L0}
L_0=\frac{3\sqrt{d_1d_2}(B_2\sqrt{d_1}+B_1\sqrt{d_2})}{2\sqrt{3}
(3-8V_*^2)(2V_*^2-3\alpha)(V_*^4-\frac{1}{2}(2\alpha+1)V_*^2+\frac{3\alpha}{8})}\;,
\end{equation}
and 
\[B_j=\frac{1}{2}\left[\frac{1}{6}\left(V^2_*-(-1)^j M_0
\right)^2-V^2_*\left(\frac{3V^2_*}{2}+1\right)-\frac{3\alpha}{2}\left(\alpha-\frac{1}{2}-2{V^2_*}\right)\right]\;,\quad j=1,2\;.\]
\end{prop}

\emph{Proof. }
Using (\ref{eqsqasy}), for $\xi \to 0$: 
\begin{equation}\label{diff_asy}
z_2-z_1=\frac{M_0\xi^2}{3}+O(\xi^4)\;,
\end{equation}
and
\begin{equation*}\label{asyfns2}
F(z_j)=
B_j\xi^4+O(\xi^6)\;,\quad j=1,2\;,
\end{equation*}
together with  (\ref{Lkn1n2}), (\ref{rad1}), (\ref{M0}),   (\ref{diff_asy}) gives $(\ref{eqLasy})$,
where 
\begin{eqnarray*}
L_0&=&\frac{M_0\sqrt{d_1d_2}}{2\sqrt{3}\Big(B_2\sqrt{d_1}-B_1\sqrt{d_2}\Big)}\;.\nonumber
\end{eqnarray*}
Multiplying the numerator and the denominator by
$B_2\sqrt{d_1}+B_1\sqrt{d_2}$
, we arrive at (\ref{L0}). 
\hfill $\Box$
 
 It is shown in Appendix I 
that $L_0$ is positive for $\alpha>0$, and $V\le C_R$.



\section{Roots  and poles of the kernel function $L(\xi)$}\label{RPLXI}
\label{dispsec}
Having obtained the expression for the kernel function $L(\xi)$, we can now derive expressions for the dispersion relations of the inhomogeneous lattice. We also show that these relations can be used to predict the behaviour of the argument of the function $L$, which is needed in the computation of the energy release rate for the propagating crack in section 7.
Let $m=1$ and let
\[p=3r(\Lambda_1^*)r(\Lambda_2^*)\sqrt{z_1^2-1}\sqrt{z_2^2-1}\]
and
\begin{equation}\label{hdelt}
 q=\frac{r(\Lambda_1^*)F(z_2)\sqrt{z_1^2-1}-r(\Lambda_2^*)F(z_1)\sqrt{z_2^2-1}}{z_2-z_1}
 \end{equation}
so that in (\ref{Lkn1n2})
\[L(\xi)=\frac{p}{q}
\;.\]
We now investigate the zeros of $p$ and $q$ of the above expression. This is carried out by setting $\varepsilon=0$ and $\Omega=\xi V$, then
 solving the equations $p=0$ and $q=0$
for $\Omega$. 

The roots of the equation 
$p=0$ 
are given by:
\begin{eqnarray}
\Omega_1^{(N)}(\xi)&=&\sqrt{6\mu_2}|\cos(\xi/4)|\;,
\label{root1}\\
\Omega_2^{(N)}(\xi)&=&[2\mu_2\sin^2(\xi/4)
+4\mu_1 \sin^2(\xi/2)
]^{1/2}\;,
\label{root2}\\
\Omega_3^{(N)}(\xi)&=&\sqrt{6\mu_2}|\sin(\xi/4)|\;,
\label{root3}\\  
\Omega_4^{(N)}(\xi)&=&[2 \mu_2 \cos^2(\xi/4)+4\mu_1 \sin^2(\xi/2)]^{1/2}\;.\label{root4}
\end{eqnarray}
Next we obtain the roots of the equation 
$q=0$.
This equation, after multiplication by 
\[r(\Lambda_2^*)F(z_2)\sqrt{z_1^2-1}+r(\Lambda_1^*)F(z_1)\sqrt{z_2^2-1}\;, \]
leads to 
\begin{equation}\label{exteq}
\frac{F(z_2)^2({z_1^2-1})-F(z_1)^2({z_2^2-1})}{(z_2-z_1)}=0\;.
\end{equation}
Solutions of this extended equation are then (\ref{root3}), (\ref{root4}) and 
\begin{eqnarray}
\Omega^{(D)}_1(\xi)&=&\sqrt{ 2\mu_1+\mu_2-\sqrt{3\mu_1^2+(\mu_1-\mu_2)^2}}|\sin(\xi/2)|\;,\label{Droot1}\\
\Omega^{(D)}_2(\xi)&=&\sqrt{6\mu_1}|\sin(\xi/2)|\;,\label{Droot2}
\\
\Omega^{(D)}_3(\xi)&=&\sqrt{ 2\mu_1+\mu_2+\sqrt{3\mu_1^2+(\mu_1-\mu_2)^2}}|\sin(\xi/2)|\;,
\label{Droot3}
\end{eqnarray}
and it remains to determine which of the roots of (\ref{exteq})  are those of (\ref{hdelt}). Note that for each dispersion relation 
$\Omega(\xi+4\pi)=\Omega(\xi)$.

Functions  (\ref{root3}) and (\ref{root4}) are roots of equation (\ref{hdelt}) and are in fact removable singularities of $L(\xi)$.

The choice of $\alpha$ and $\xi$ determines when  (\ref{Droot1})--(\ref{Droot3}) are solutions of $q=0$. Here, we present the results for the case $\alpha=1$ and give the description of the behaviour of the argument of $L$. The function $\Omega_1^{(D)}$  is a solution of (\ref{hdelt}). When
\begin{equation}\label{eqreg1}
0.6667\le \xi/\pi\le  1.3333\quad  \text{ and } \quad 2.6667\le \xi/\pi\le  3.3333, 
\end{equation}
we have $\Omega_2^{(D)}$ is a solution of $q=0$. For $\xi/\pi$ 
satisfying \begin{equation}\label{eqreg3}
0.7614 \le \xi/\pi\le 1.2386\quad \text{ and } \quad 2.7614 \le \xi/\pi\le 3.2386,
\end{equation}
 $\Omega_3^{(D)}$ is a solution of $q=0$. The inequalities  (\ref{eqreg1}) and (\ref{eqreg3})  and their derivations
 are   discussed in the Appendix II.


In Figure
\ref{figdisp3}(a) we plot the  dispersion relations (\ref{root1})--(\ref{root4}) and (\ref{Droot1})--(\ref{Droot3}), as functions of the wavenumber $\xi$, for $\mu_1=\mu_2=1$.
 Included in this figure is the ray $\Omega=V \xi$ corresponding to the speed $V=0.4504$. By comparison with the diagram of $\text{arg}(L(\xi))$ in Figure \ref{figdisp3}(b), we see each intersection of the ray $\xi V$ with the dispersion curves corresponds  to a jump in $\text{arg}(L(\xi))$.
 
 At approximately $\xi=\pi$, we have an additional jump in $\text{arg}(L(\xi))$. This is a result of the behaviour of the factor $z_2-z_1$ contained in $L$ (see (\ref{Lkn1n2}) and (\ref{biquadroots})). This factor, as shown in Figure \ref{figdisp3}(c), is either real or purely imaginary. As we approach $\xi=\pi$, this factor goes moves from the negative imaginary axis in the complex plane, to the positive  real axis (note that $z_2-z_1$ is the radical function of (\ref{eqzj})
 where  the positive square root is taken). The corresponding effect is a jump in $\text{arg}(L(\xi))$ of $\pi/2$. 
 
 For a lower crack speed, $V=0.21395$, we present the dispersion diagram along with the  ray $\xi V$ in Figure \ref{figdisp3c}(a). On the corresponding picture for $\text{arg}(L(\xi))$ in Figure \ref{figdisp3c}(b), we see this function jumps when the ray intersects the dispersion curves. In particular, the  ray  $\xi V$ intersects the curve corresponding to $\Omega_3^{(D)}$  in  the second region of  (\ref{eqreg3}), which again results in a change in the value of  $\text{arg}(L(\xi))$. Also, 
 in Figure \ref{figdisp3c}(c), at approximately, $\xi=3.4\pi$, we observe that the factor $z_2-z_1$ moves from the positive real axis to the positive imaginary axis. Again, the result we see is a jump of $\pi/2$ in $\text{arg}(L(\xi))$ at this point. 
 
 Finally, for $\mu_1=1$ and $\mu_2=10$, $(\alpha=0.1)$, $V=0.799076$ we present the dispersion relations, the plot of $\text{arg}(L(\xi))$ and the imaginary part of the factor $z_2-z_1$, in Figure \ref{figdisp3e}(a)--(c). Comparing with Figure  \ref{figdisp3}, we see the distance between the dispersion curves along with their height has increased, which can lead to greater absolute value of the area under the curve traced by the argument of $L$.
 \begin{figure}[htbp]

\begin{center}
\hspace{-0.75in}\includegraphics[width=1.1\textwidth]{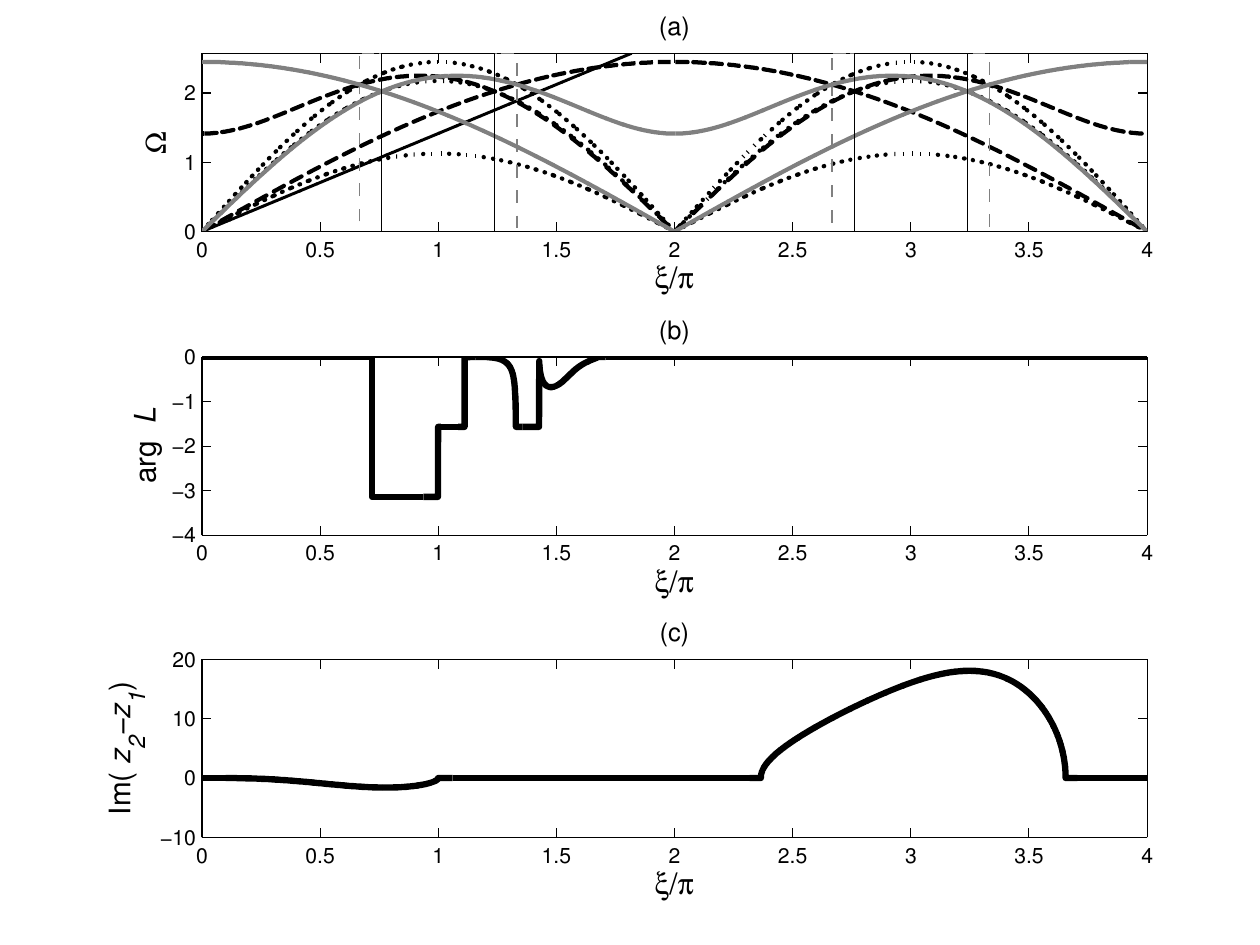}
\end{center}


\caption{For $\mu_1=\mu_2=1$, $\alpha=1$,  figure (a)  shows the  dispersion relations (\ref{root1})--(\ref{root4}) and (\ref{Droot1})--(\ref{Droot3}), along with the  ray $\xi V$ for $V=0.4504$. The roots of $L$ are given by the solid curves, poles are shown by  curves with the small dashes and the removable singularities are given by the dashed  curves. The regions given in (\ref{eqreg1}) and (\ref{eqreg3}) where $\Omega^{(D)}_2$ and $\Omega^{(D)}_3$ are roots of $q=0$ are marked by the dashed and solid vertical lines. For comparison, we give the plot of (b) $\text{arg}(L(\xi))$, and (c)  the imaginary part of the factor $z_2-z_1$ contained in the representation (\ref{Lkn1n2}) of $L$. }
\label{figdisp3}

\end{figure}






 \begin{figure}[htbp]

\begin{center}
\hspace{-0.65in}\includegraphics[width=1.1\textwidth]{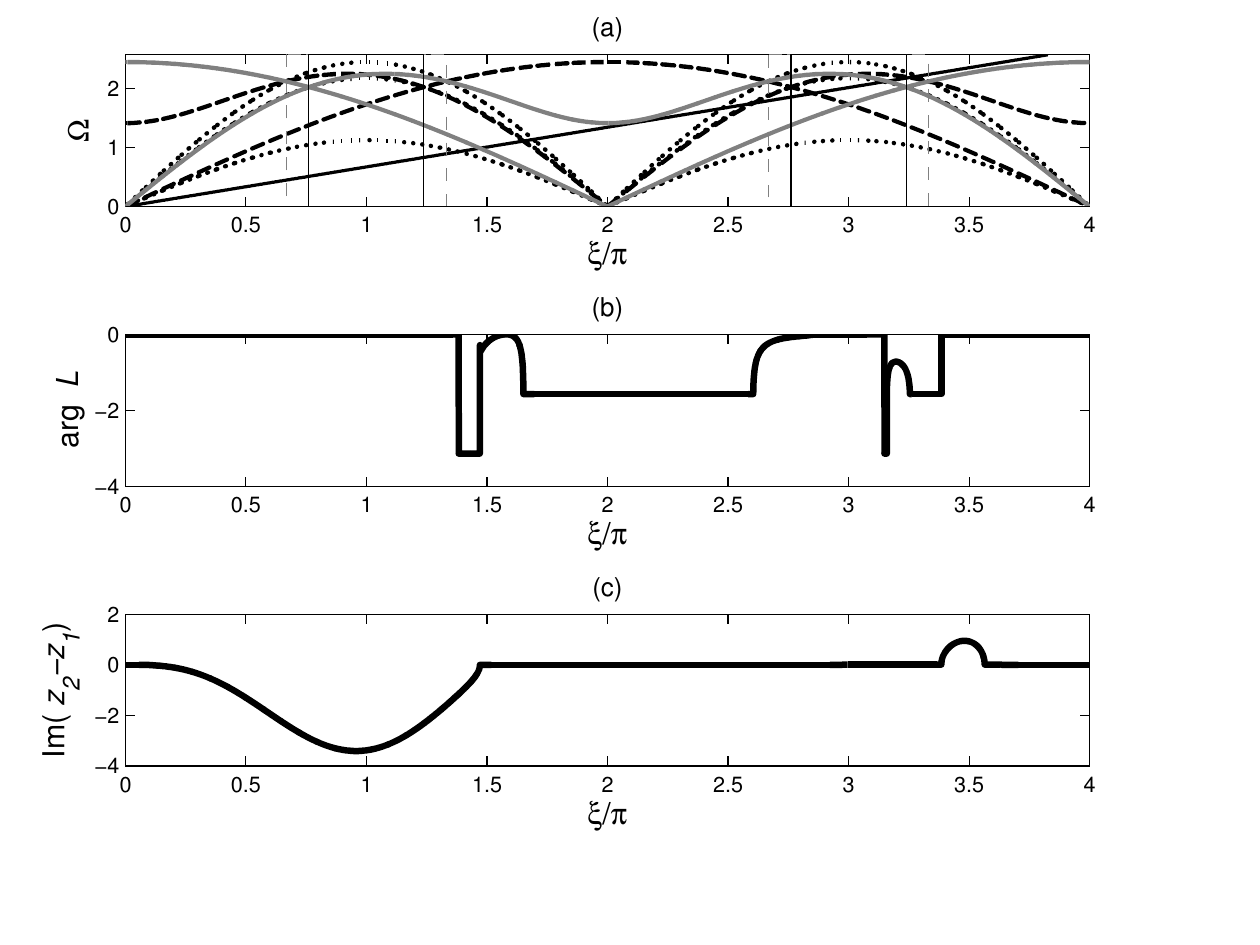}
\end{center}


\caption{For $\mu_1=\mu_2=1$, $\alpha=1$, in figure (a) we show the  dispersion relations (\ref{root1})--(\ref{root4}) and (\ref{Droot1})--(\ref{Droot3}), along with the  ray $\xi V$ for $V=0.21395$. 
The description of the curves in the dispersion diagram is the same as in Figure \ref{figdisp3}. 
We again  give the plot of (b) $\text{arg}(L(\xi))$ and  (c) the function Im($z_2-z_1$), for the purpose of comparison.}
\label{figdisp3c}

\end{figure}






\begin{figure}[htbp]

\begin{center}
\hspace{-0.75in}\includegraphics[width=1.1\textwidth]{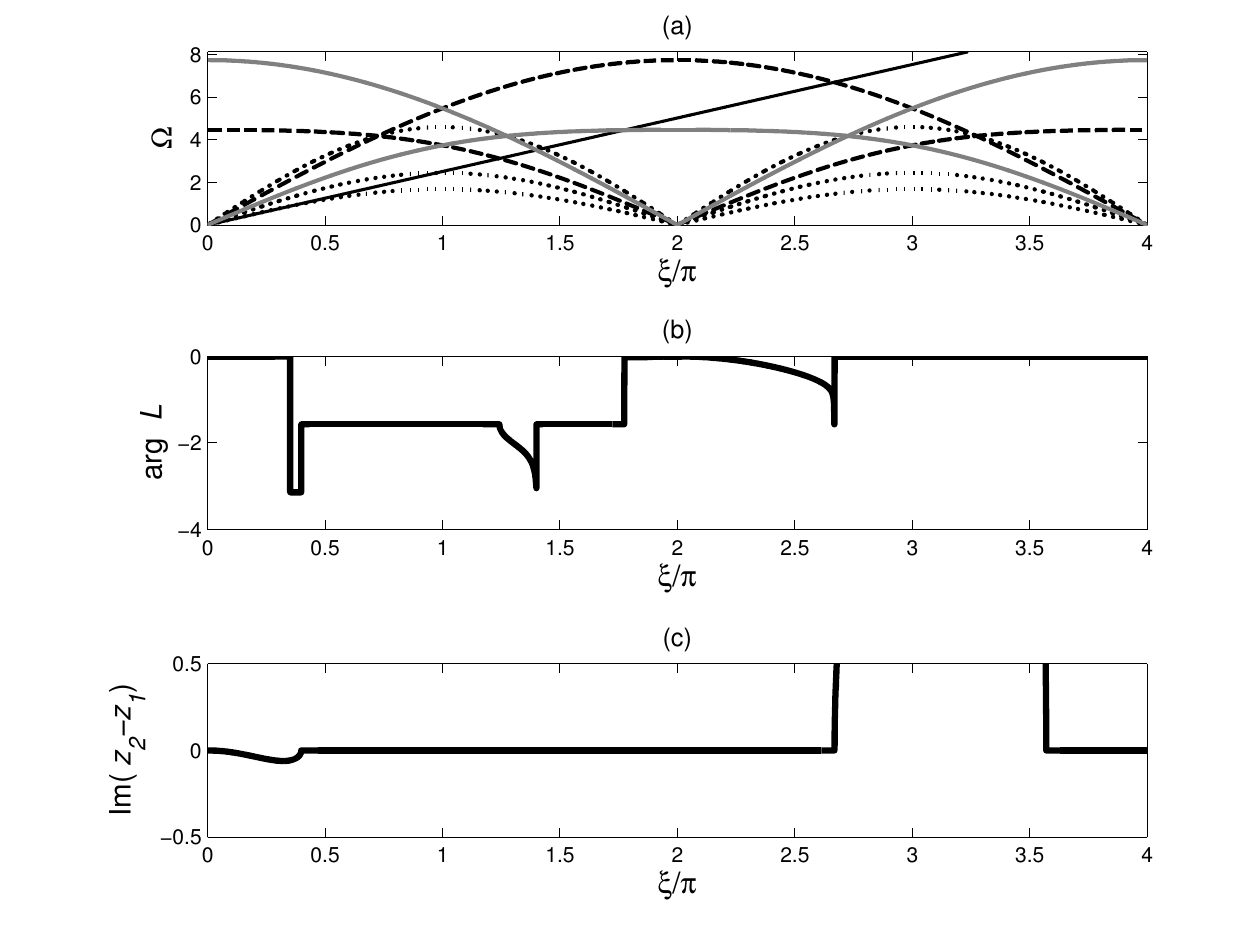}
\end{center}


\caption{(a) For $\mu_1=1$, $\mu_2=10$, $\alpha=0.1$, we plot the  dispersion relations (\ref{root1})--(\ref{root4}) and (\ref{Droot1})--(\ref{Droot3}), along with the  ray $\xi V$ for $V=0.79976$. 
The description of the dispersion curves is given in Figure \ref{figdisp3}.
The plots of (b) $\text{arg}(L(\xi))$ and (c) $\text{Im}(z_2-z_1)$ are also presented. }
\label{figdisp3e}

\end{figure}

\section{Solution of the Wiener-Hopf equation}\label{SOLWH}
\label{secWHE}
\subsection{Factorisation of the Wiener-Hopf equation (\ref{eqWH})}
Here, we use the solution presented in \cite{Slepyan} and extend it to the situation when the bonds within the lattice have a contrast in stiffness in the principal lattice directions. 
We define the class $\CJ$ as the set of complex valued functions $f:\mathbb{R}\to \mathbb{C}$ satisfying
\[{\rm Re}(f(\xi))={\rm Re}(f(-\xi))\;, \quad {\rm Im}(f(\xi))=-{\rm Im}(f(-\xi))\;.\] 
For $\varepsilon>0$,   the kernel $L\in \mathcal{J}$ 
 and  satisfies 
\begin{equation}
\begin{array}{l}
\displaystyle{{\rm Ind}(L(\xi))=\frac{1}{2\pi}[{\rm arg}(L(+\infty))-{\rm arg}(L(-\infty))]=0}\;,\\ \\
{\rm arg}(L(\xi))=-{\rm arg}(L(-\xi)) \quad \text{ and }\quad   L(\pm \infty)=1\;, 
\end{array}\label{propL1}
\end{equation}
where $\text{Ind}(L(\xi))$ represents the number of times the contour traced by $L(\xi)$ in the complex plane winds around the origin.
The proof of these properties are given in Appendix III.

Then, the above conditions  allow $L(\xi)$ to be factorised in the form
\begin{equation*}\label{factL}
L=L_+L_-\;,
\end{equation*}
where 
\begin{equation}\label{Cauchyint}
L_\pm(\varepsilon+\text{i}\xi V, \xi)=\text{exp}\left(\pm\frac{1}{2\pi \text{i}} \int ^\infty_{-\infty}\frac{\ln L(\varepsilon+\text{i} s V, s)}{s-\xi} d s\right)
\end{equation}
with $\varepsilon \to +0$, $\text{Im } \xi>0$ for $L_+$ and $\text{Im }\xi<0$ for $L_-$.

 The function $L_+$ ($L_-$) is an analytic function of $\xi$ in the upper (lower) half-plane. 
 The introduction of the small positive parameter $\varepsilon$ into $L$, shifts its zeros and poles (located on the real axis) into the upper or lower half of the complex plane. Following this shift, the dispersion curves, discussed in section \ref{RPLXI}, can be then used to identify in which half plane each root and pole is located 
 (see Slepyan \cite[Chapter 2]{Slepyan}).

Suppose $\xi_z$ and $\xi_p$ are wavenumbers which correspond to the intersection of $\Omega=\xi V$ with the zeros and poles of $L$, respectively (see (\ref{root1})--(\ref{root4}) and (\ref{Droot1})--(\ref{Droot3})), on the dispersion diagram. Also at these intersection points, let $V<V_g$, where $V_g=d\Omega/d\xi$ is the group velocity. Then, in this case $\xi_z$ and $\xi_p$ will be located in the lower half plane after the introduction of $\varepsilon$ in $L$. Similarly, if $V>V_g$ then $\xi_z$ and $\xi_p$ are shifted into the upper half of the complex plane (see \cite[Section 3.3.3]{Slepyan}). From the definition (\ref{Cauchyint}), it follows that  $L_+$ $(L_-)$
contains all singular and zero points of $L$ which are located in the  
lower
 (upper) half of the complex plane for $\varepsilon>0$.      
 
  Using (\ref{factL}),  equation (\ref{eqWH}) can then be written as 
\begin{equation}
\label{eqfactored}
L_+ \mathcal{S}_++\frac{\mathcal{S}_-}{L_-}=\left(\frac{1}{L_-}-L_+\right) \phi^F\;,
\end{equation}
which is the same as (12.58) of \cite{Slepyan}.

Here, the left-hand side of (\ref{eqfactored}) is the sum of two analytic functions, one analytic in the upper half of the complex plane, the other analytic in the lower half of the complex plane and in order to solve the problem, we must separate the terms in the right-hand side in the same way. This is carried out by introducing the load $\phi$ in such a way that it allows for this additive split.
In Appendix IV, an outline for this procedure  is presented for the case when the load $\phi$ is chosen so that  the right-hand side can be represented as a linear combination of Dirac delta functions.


In what follows, we will consider the situation 
when the  load  applied along the crack faces 
which generates  term of the form $\mathcal{C}\delta(\xi)$, in the right-hand side of (\ref{eqfactored}) (see Appendix IV for the derivation). Here, $\mathcal{C}$ is a constant which represents the intensity of the load. 
 Then $\mathcal{S}_+$ and $\mathcal{S}_-$ have the form
\begin{equation}\label{Qpma}
\begin{array}{c}
\displaystyle{\mathcal{S}_+=
  \frac{\CC}{L_+(\xi)(\varepsilon-{\rm i}\xi)}\;,}\qquad 
\displaystyle{\mathcal{S}_-=  
\frac{L_-(\xi)\,\CC}{\varepsilon+{\rm i}\xi}\;,\quad \varepsilon \to +0\;,}
\end{array}
\end{equation}
\section{Energy release rate ratio}\label{ERR_sec}

Here, 
we investigate the dependence of energy release  rate ratio \cite{Slepyan} on the stiffness contrast parameter $\alpha$. This is defined as
\begin{equation}\label{ERRexp}
\frac{G_0}{G}=\text{exp}\left(\frac{2}{\pi} \int^\infty_0 \frac{\text{arg}(L(\varepsilon+\text{i}s V,
s))}{s} ds\right)\;,\quad \varepsilon \to +0\;,
\end{equation}
where $G_0$ is the local energy release rate for the semi-infinite crack propagating through the lattice with speed $V$ for sub-Rayleigh wave speed and $G$ is global energy release rate for the crack propagating through the corresponding homogenised medium. This ratio describes the dissipation of energy created by breaking bonds at the crack front. 

Next we show
 (\ref{ERRexp}) can be obtained from (\ref{Qpma}).
However, we note that the derivation of   (\ref{ERRexp}) does not depend on the solution of the Wiener-Hopf equation in the previous section. 
The definitions of $G$ and $G_0$ follow \cite[Chapter 12]{Slepyan} as the energy release rates globally and locally respectively. In particular, 
as in 
\cite{Slepyan}, $G$ can be defined as a limit
\begin{equation*}\label{G}
G=\lim_{k \to 0} \varepsilon^2 \mathcal{S}_-({\rm i}k)\mathcal{S}_+({\rm i}k)\;,
\end{equation*}
and  the local energy release rate is given as 
\begin{equation*}
\label{localERR1}
G_0=\lim_{k \to +\infty} \varepsilon^2 \mathcal{S}_-({\rm i}k)\mathcal{S}_+({\rm i}k)\;.
\end{equation*}
The global energy release rate $G$ follows from the asymptotes of $\mathcal{S}_{\pm}$ as $\xi\to 0$. The asymptotes of $L_\pm$ are 
\begin{equation}\label{asyLpmk0}
\begin{array}{c}
\displaystyle{L_+(\varepsilon+{\rm i} \xi V, \xi) \sim \frac{\sqrt{L_0}}{\sqrt{\varepsilon-{\rm i}\xi}}\,\text{exp}\left(\frac{1}{\pi}\int^\infty_0 \frac{\text{arg}( L(\varepsilon+{\rm i} s V, s))}{s}\, ds\right)\;,}\\
\\
\displaystyle{L_-(\varepsilon+{\rm i} \xi V, \xi) \sim \frac{\sqrt{L_0} }{\sqrt{\varepsilon+{\rm i}\xi}}\,\text{exp}\left(-\frac{1}{\pi}\int^\infty_0 \frac{\text{arg}( L(\varepsilon+{\rm i} s V, s))}{s}\, ds\right)\;,}\end{array}
\end{equation}
for $\varepsilon\to +0$, and their derivations are found in Appendix V.

Then, owing to (\ref{Qpma}) and (\ref{asyLpmk0}),
\begin{equation}\label{tenfor}
\mathcal{S}_+ \sim \frac{\CC}{\sqrt{L_0}\sqrt{\varepsilon-{\rm i} \xi}}\,\text{exp}\left(-\frac{1}{\pi}\int^\infty_0 \frac{\text{arg}( L(\varepsilon+{\rm i} s V, s))}{s}\, ds\right)\;,
\end{equation}
\[\mathcal{S}_-\sim \frac{\sqrt{L_0} \CC}{(\varepsilon+{\rm i}\xi)^{3/2}}\,\text{exp}\left(-\frac{1}{\pi}\int^\infty_0 \frac{\text{arg}( L(\varepsilon+{\rm i} s V, s))}{s}\, ds\right)\;, \]
where $\xi \to 0$, $\varepsilon\to +0$,  and then 
\begin{equation}\label{GlobalERR}
G= \CC^2\text{exp}\left(-\frac{2}{\pi}\int^\infty_0 \frac{\text{arg}( L(\varepsilon+{\rm i} s V, s))}{s}\, ds\right) \;,\quad \varepsilon \to +0\;.
\end{equation}
The local energy release rate is then given by the asymptotes of $\mathcal{S}_\pm$ when $\xi\to \infty$. As a result of 
(\ref{propL1}) 
and (\ref{Cauchyint}), $L_\pm\to 1$ as $\xi\to \infty$.  In the same limit we have 
\begin{equation*}
\mathcal{S}_+ \sim \frac{\CC}{{\varepsilon-{\rm i} \xi}}\;,\quad \mathcal{S}_-\sim \frac{\CC}{\varepsilon+{\rm i}\xi} \;,\quad \xi\to \infty\;, \varepsilon\to +0\;,
\end{equation*}
and so $G_0=\CC^2$. Then this and  (\ref{GlobalERR}) imply (\ref{ERRexp}).
\begin{figure}[htbp]
\centering

{\includegraphics[width=0.8\textwidth]{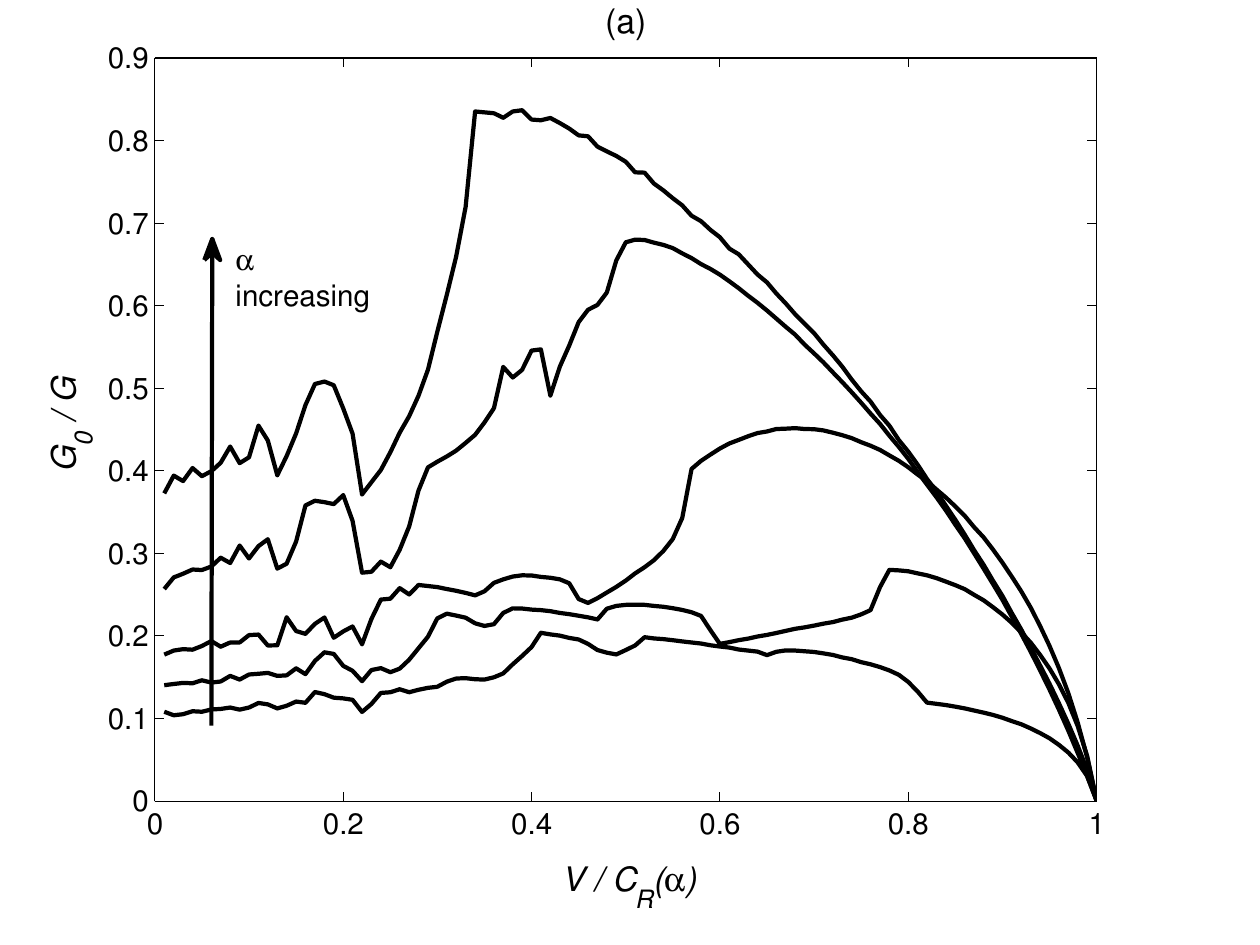}}
{\includegraphics[width=0.8\textwidth]{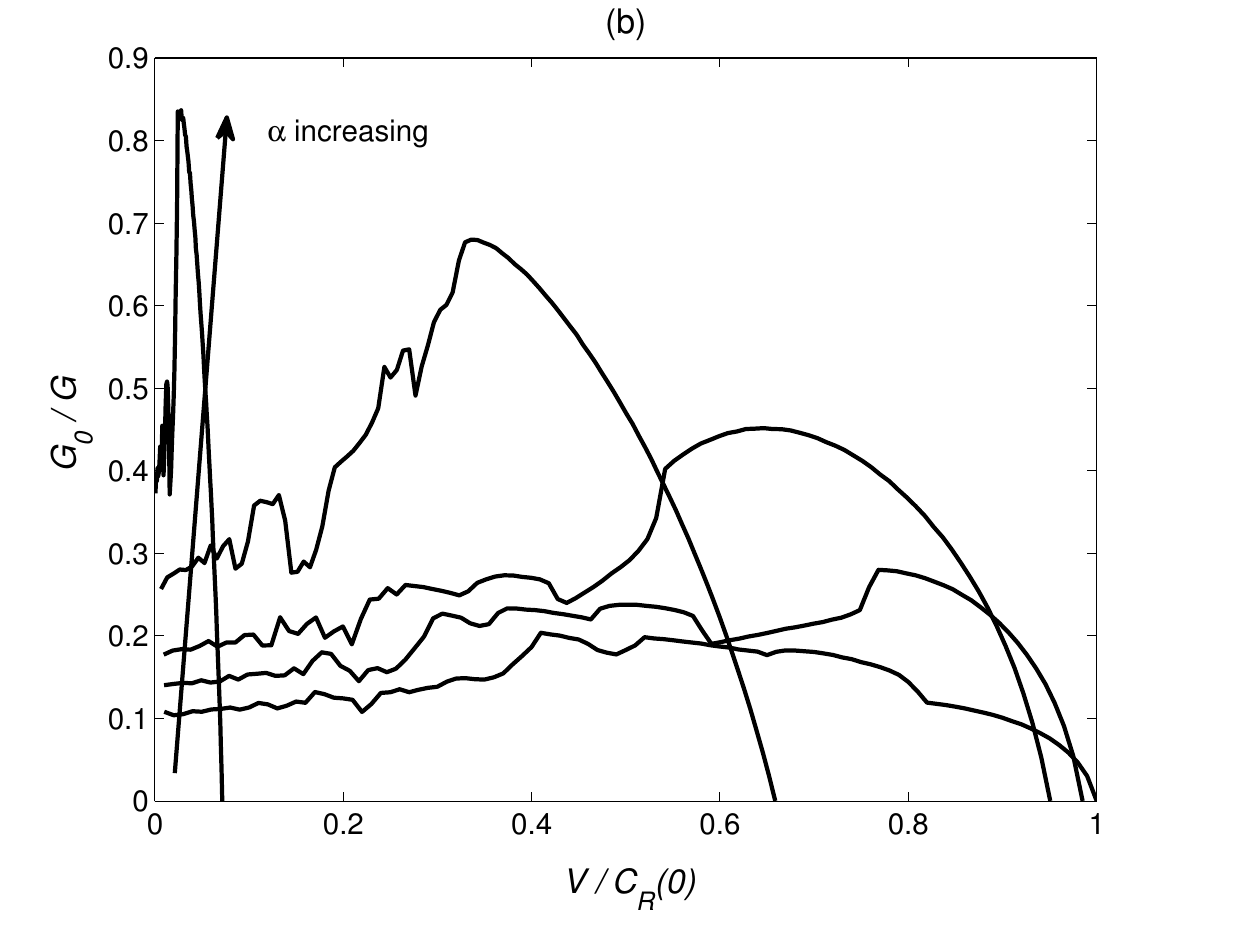}}
\caption{(a): The energy release rate ratio as a function of the normalised speed $V/C_R(\alpha)$, based on formula (\ref{ERRexp}). (b) The ratio $G_0/G$ as a function of the crack speed normalised by the supremum of  $C_R(\alpha)$ over $\alpha>0$ ($C_R(0)=\sqrt{3}/2$). We show the behaviour of this ratio for the stiffness contrasts $\alpha=0.05, 0.1,0.2, 1$, and $100$.}
\label{ERR_ratio_fig_2}
\end{figure}

\subsection{Sensitivity of the energy release rate ratio to the stiffness contrast}
In this section, we set $m=\mu_1=1$, so that
  $\alpha=1/\mu_2$, and we investigate the behaviour of the energy release rate ratio (\ref{ERRexp}) as a function of crack speed, when  the stiffness ratio is increased (which corresponds to the decrease in the stiffness of the inclined bars). The critical crack speed $C_R(\alpha)$, in this example, has the form (see (\ref{subR})):
\begin{equation}\label{critspeed}
C_R(\alpha)\Big|_{m=1, \mu_1=1}=\frac{1}{2\alpha^{1/2}}\sqrt{ 2\alpha+1-\sqrt{3\alpha^2+(\alpha-1)^2}}\;.
\end{equation}
The lattice Rayleigh speed $C_R(\alpha)$ is decreasing monotonically as $\alpha$ increases. The upper bound of this function is $C_R(0)=\sqrt{3}/2$, (when the inclined bars are rigid) and for $\alpha$ tending to infinity (the inclined  bars are becoming much more softer compared to the horizontal bars) the speed $C_R(\alpha)$  tends to zero.
\begin{figure}[htbp]
\centering
\hspace{-0.1in}\includegraphics[width=0.8\textwidth]{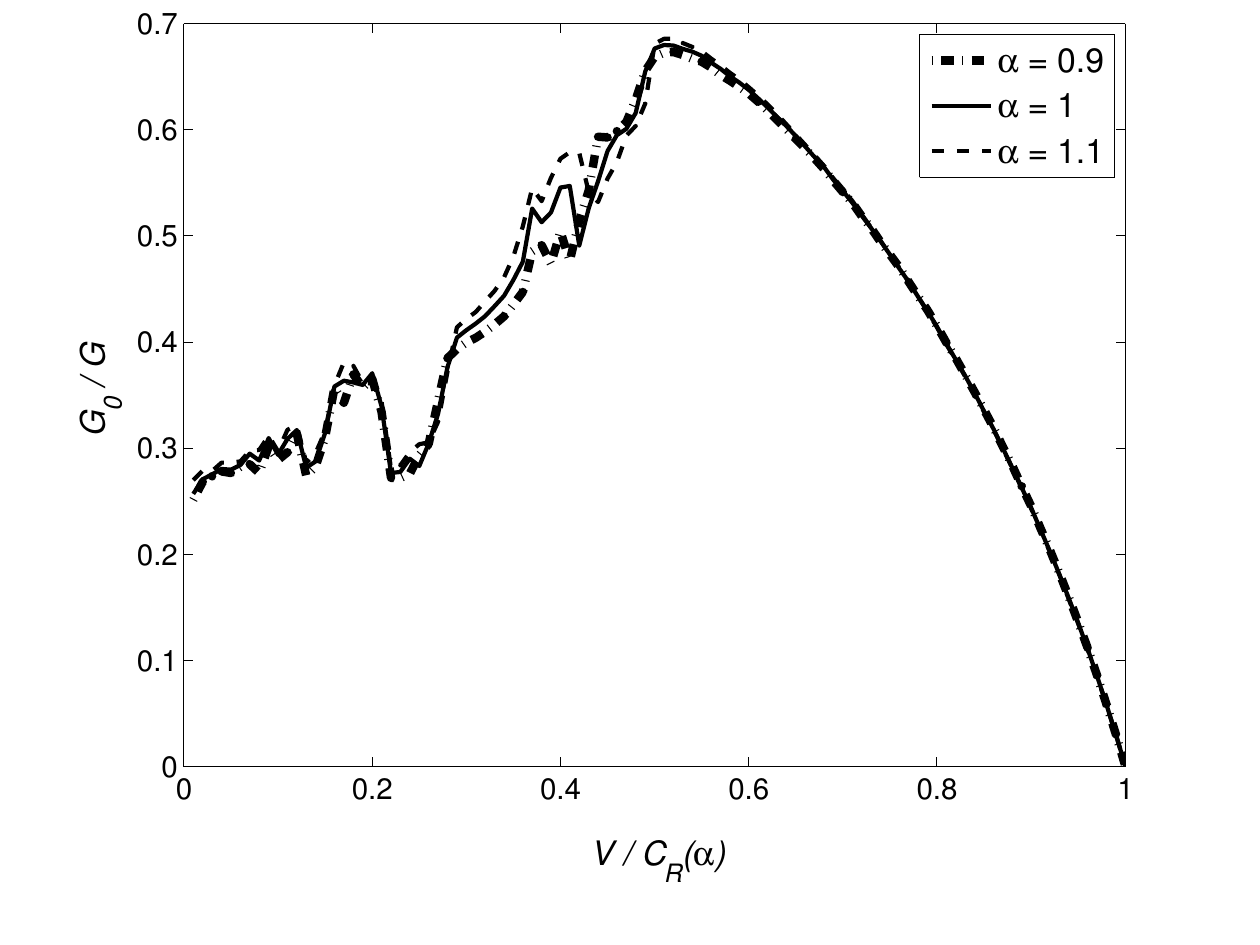}
\caption{The energy release rate ratio as a function of the normalised speed $V/C_R(\alpha)$,  computed using formula (\ref{ERRexp}), for the values $\alpha=0.9, 1$ and $1.1$.}
\label{ERR_ratio_fig_3}
\end{figure}
In Figure \ref{ERR_ratio_fig_2}(a), we plot the energy release rate ratio as a function of normalised crack speed $V/C_R(\alpha)$, for $\alpha=0.05, 0.1, 0.2, 1, 100$. 
 The crack, for low speeds, does not propagate steadily. The intrinsic feature of the model discussed here is that the crack propagates uniformly with speed $V$. If this assumption is not satisfied then the model  does not describe accurately the propagation of the crack, which is depicted by the non-smooth part of the curves on Figure \ref{ERR_ratio_fig_2}(a), e.g. for approximately $V/C_R(\alpha)< 0.5$ when $\alpha=1$. This region of instability is 
 caused by the highly  oscillatory behaviour of $\text{arg}(L(\xi))$ for low crack speeds, in (\ref{ERRexp}), as observed in section \ref{RPLXI}. From Figure \ref{ERR_ratio_fig_2}(a), we see the percentage of speeds less then $C_R(\alpha)$, where this region is located, is decreasing as $\alpha$ is increasing.
 
 Contained in Figure \ref{ERR_ratio_fig_2}(b), is the plot of the energy release rate ratio  against the crack speed which has been normalised by $C_R(0)=\sqrt{3}/2$.
  For $\alpha\ge 0.05$ the ratio $G_0/G$ tends to zero as we approach critical speed $C_R(\alpha)$ predicted by (\ref{critspeed}).
For $\alpha>0.1$, in the region of instability, the ratio $G_0/G$ 
increases until it reaches  a speed where there is a global maximum. For speeds greater than this, we observe the ratio follows a smooth curve as it decreases to zero at the critical Rayleigh speed (\ref{critspeed}). In this region, this behaviour shows that the model describes the  steady propagation of the semi-infinite crack through the inhomogeneous lattice.


In \cite{Mishetal}, the dynamic problem for the propagation of a crack situated in a square-cell lattice, whose rows contain particles of contrasting 
mass, 
 is discussed.  The sensitivity to the contrast in mass, of the ratio of the local energy release rate $G_0$ (for the lattice) to the global energy release rate  $G$ (for the corresponding homogenised material) was also investigated. It was seen the values of $G_0/G$, for high crack speeds, increased monotonically as the contrast in mass was reduced. Here, as shown in Figures \ref{ERR_ratio_fig_2}(a), (b) and \ref{ERR_ratio_fig_3}, we do not have any  monotonicity in the behaviour of $G_0/G$ for different $\alpha$.  
 For example, in Figure \ref{ERR_ratio_fig_2}(a), the curve for $\alpha=0.2$ intersects those for $\alpha=1, 100$ at approximately $V/C_R(\alpha)=0.83$ (where we are in the region of stability for all these curves). For speeds higher than this value, the ratio of $G_0/G$ for $\alpha=0.2$ is greater than that for $\alpha=1$ and $\alpha=100$. A similar example can be seen for the curve corresponding to $\alpha=0.1$.
 
 
 It was also seen in \cite{Mishetal}, for low crack speeds corresponding to the instability region of the model, the values of $G_0/G$ where not monotonic as a function of the mass contrast parameter. A similar feature is also observed here for low crack speeds as the stiffness contrast parameter $\alpha$ is varied.
 In Figure  \ref{ERR_ratio_fig_3}, we plot the energy release rate ratio for $\alpha=0.9,1$ and $1.1$. 
  For speeds $V/C_R(\alpha)<0.5$ (the region of instability), there exists
   regions where all curves are intersecting one another and overlapping. For instance,  when $0.35<V/C_R(\alpha)<0.5$, 
 again  there  is no monotonicity in the behaviour of $G_0/G$ as 
   $\alpha$ varies.
  We note that the critical speed in the scalar case \cite{Mishetal}, was a material constant and did not depend on the contrast ratio.

 

\section{Stress intensity factor in the homogenisation approximation}\label{SIF_hom_approx}

In this section, we derive the expression for the Mode I stress intensity factor for the semi-infinite crack in the case when the  load $\phi$
applied along the crack faces
 generates the term $\CC \delta (\xi)$ in the right-hand side of (\ref{eqfactored}), with $\CC$ being the load intensity. The solution  of this problem is given in  (\ref{Qpma}).

\begin{figure}[htbp]
\centering
\includegraphics[width=0.8\textwidth]{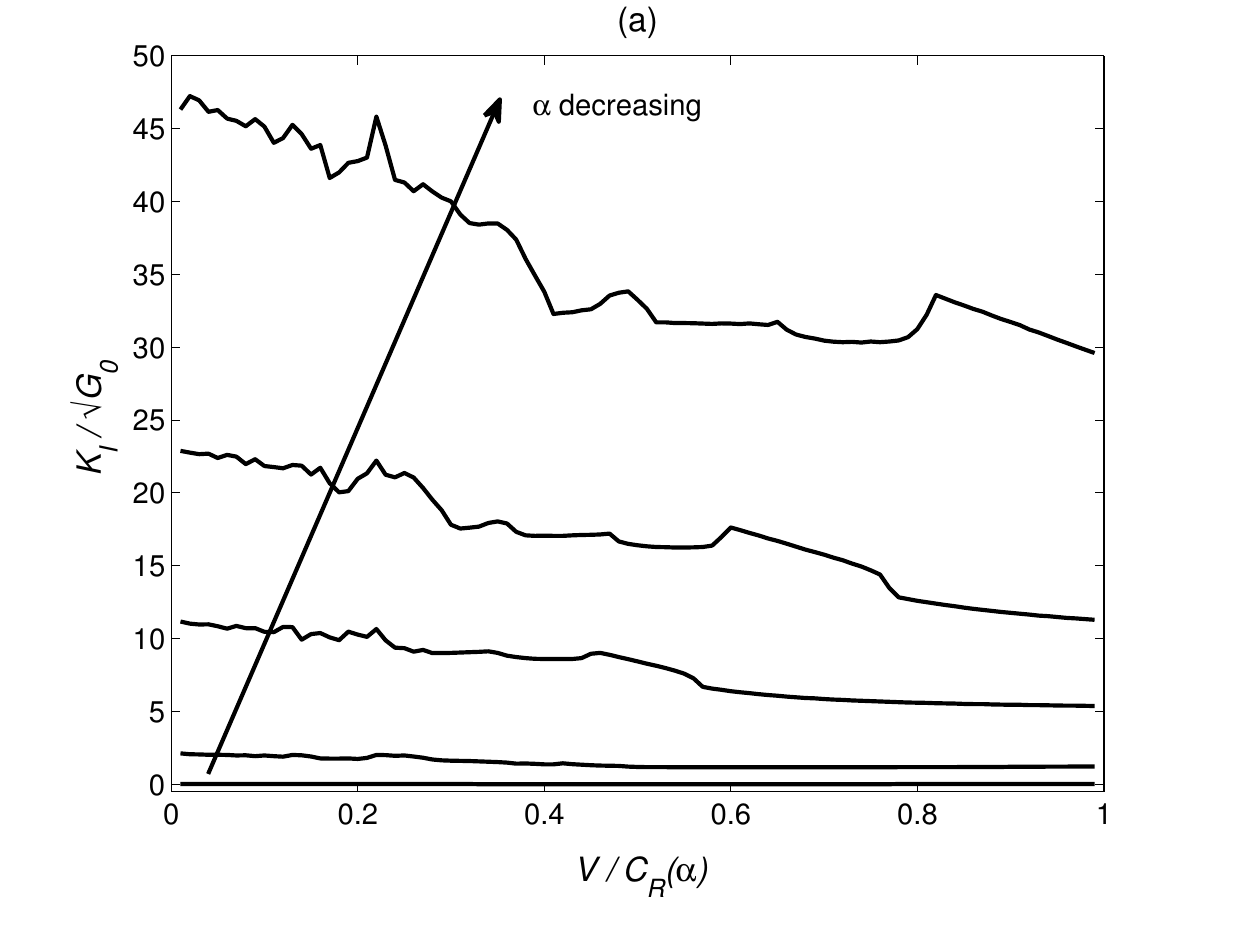}

\includegraphics[width=0.8\textwidth]{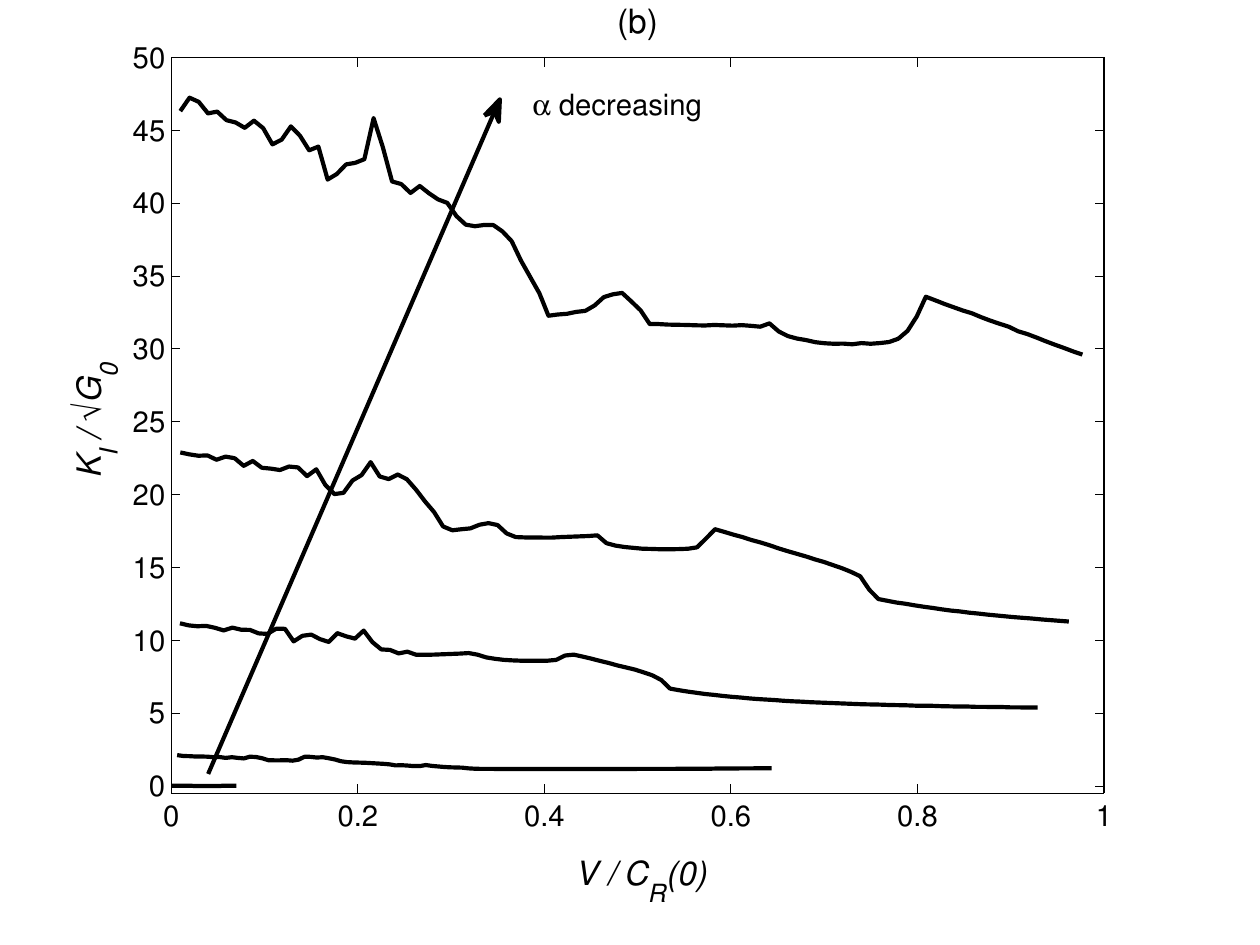}
\caption{The normalised stress intensity factor ${K}_I/\sqrt{G_0}$ as a function of  (a) the normalised crack speed $V/C_R(\alpha)$ and (b) the normalised crack speed $V/C_R(0)$, based on formula (\ref{Qahead}), for $\alpha=0.05, 0.1, 0.2, 1, 100$. Here $C_R(0)=\sqrt{3}/2$ is the supremum of (\ref{critspeed}) for $\alpha>0$.}
\label{SIF}
\end{figure}
Using the fact that the energy required to break the crack front bond $G_0=\CC^2$ (see the previous section),  we compute the inverse Fourier transform of $\mathcal{S}_+$ in (\ref{tenfor}), so that to leading order 
\begin{equation}\label{Qahead}
\mathcal{S}(\eta)\sim \sqrt{\frac{G_0}{L_0 \pi \eta}}
\,\text{exp}\left(-\frac{1}{\pi}\int^\infty_0 \frac{\text{arg}( L(\varepsilon+{\rm i} s V, s))}{s}\, ds\right)\;, \quad  \text{ for }\eta>0
\end{equation}
which depends on $\alpha$ and here $L_0$ is given in Proposition \ref{Propasy0}. 
 Here, the expression $\mu_2 \mathcal{S}(\eta)$ represents the tensile force in the inclined bonds ahead of the crack.
Let the normal traction ahead of  the crack be $\sigma$, then 
\begin{equation}\label{ffstress}
\sigma \sim
 \mu_2\sqrt{\frac{3G_0}{L_0 \pi \eta}}
\,\text{exp}\left(-\frac{1}{\pi}\int^\infty_0 \frac{\text{arg}( L(\varepsilon+{\rm i} s V, s))}{s}\, ds\right) \;.
\end{equation}
The Mode I stress intensity factor $K_I$, in the far-field is defined by the formula
\begin{equation}\label{nfstress}
K_I=\lim_{\eta\to +0}\sqrt{2\pi \eta}\,\sigma\;.
\end{equation}
On comparison of (\ref{ffstress}) and (\ref{nfstress}), we have the stress intensity factor ${K}_I$ is 
\begin{equation}\label{SIFex}
K_I(\alpha)=\mu_2  \sqrt{\frac{6G_0}{L_0 }}
\,\text{exp}\left(-\frac{1}{\pi}\int^\infty_0 \frac{\text{arg}( L(\varepsilon+{\rm i} s V, s))}{s}\, ds\right)\;.
\end{equation}

Now, setting $\mu_1=1$ and allowing $\mu_2$  to vary, we investigate the behaviour of $K_I$ as a function of $V$ and $\alpha$. 
The expression for $K_I/\sqrt{G_0}$ is given as a function of the normalised speed $V/C_R(\alpha)$ in Figure \ref{SIF}(a) and as a function of the  normalised speed $V/C_R(0)$  in Figure \ref{SIF}(b) for $\alpha=0.05, 0.1, 0.2, 1, 100$ (where $V\le C_R(\alpha)$, see (\ref{critspeed})).

 As discussed in the previous section, at low speeds the model presented here does not describe the propagation of the crack through the inhomogeneous lattice accurately. This is due to the highly oscillatory behaviour of the argument of $L(\xi)$ and it leads to regions of instability, represented by the non-smooth behaviour in $G_0/G$ for low speeds as seen in Figure \ref{ERR_ratio_fig_2}. Since the expression $K_I/\sqrt{G_0}$  contains the ratio 
 $\sqrt{G/G_0}$  
  (see (\ref{ERRexp}) and (\ref{SIFex})), we also see a similar behaviour in $K_I/\sqrt{G_0}$ in Figure \ref{SIF}(a) and (b). 
  For those speeds in this unstable region there is little physical significance of the results presented in the figures below. As we increase the speed of the crack, we leave the region of instability, and the stress intensity factor becomes a smooth function of the crack speed, and tends to a finite value for $V$ tending to the Rayleigh speed $C_R(\alpha)$ in (\ref{critspeed}), as shown in Figure \ref{SIF}(a). For $\alpha<1$, in the vicinity of $C_R(\alpha)$, $K_I$ is a monotonically decreasing function for increasing $V$. When $\alpha\ge 1$, outside the instability region, $K_I$ passes through a global minimum before the converging to a finite value at $C_R(\alpha)$.  From the figure, we can see that  overall the behaviour of the stress intensity factor is increasing as we decrease the stiffness contrast $\alpha$.


\section{Conclusions}\label{Conclusions1}

The paper has brought together analytical insight on a propagating semi-infinite fault in an elastic lattice with the computational experiments and physical interpretation of the  fields in such a lattice structure.
One of the important issues discussed here is the crack stability for different regimes of the crack speed and different values of elastic parameters of the anisotropic lattice. 

Although it appears that the analytical solution has a serious limitation due to the assumption of the steady crack propagation, it has been shown in  \cite{knife} that in the non-steady regime the averaging procedure appears to be viable. In this case, the averaged solution follows the prediction of the linear model constructed for the case of the steady crack propagation through the lattice.  

It is also noted that the low-speed unsteady regime, predicted by the model, appears to be consistent with the results obtained by other approaches (see, for example, \cite{MarderGross}), and it has a clear physical interpretation of an accelerating crack at the initial stage (low speed) of the crack advance. 

In particular, the crack stability for different values of the lattice contrast is a challenging issue, which has been comprehensively addressed here.

\vspace{0.2in} \bf Acknowledgments. \rm M. J. Nieves would like to acknowledge the financial support of the U.K. Engineering and Physical Sciences Research Council through the research grant   EP/H018514/1.
Support through European grant   FP7-PEOPLE-IAPP 
 (PIAPP-GA-284544-PARM-2), held by G.S. Mishuris, is also gratefully acknowledged. 


\renewcommand{\theequation}{I-\arabic{equation}}    
  
  \setcounter{equation}{0}  
 \section*{Appendix I: The sign of $L_0$}
 
 In this section we show that for $V\le C_R$ and $\alpha>0$ the constant $L_0$ in (\ref{L0}) of Proposition \ref{Propasy0}, section \ref{WHeqcrack}, is positive.
 
Note that by Proposition \ref{prog}, for $V_*^2 \le C_R^2m/\mu_2$
\begin{equation}\label{pos1}
3-8V_*^2>0\;,\quad 2V_*^2-3\alpha<0 
\end{equation}
and
\begin{equation}\label{pos2}
V_*^4-\frac{1}{2}(2\alpha+1)V_*^2+\frac{3\alpha}{8}> 0\;.
\end{equation}
Also 
\begin{equation}\label{pos3}
\sqrt{d_1d_2}=8\sqrt{3}\sqrt{V_*^4-\frac{2\alpha+1}{2}V_*^2+\frac{3}{8}\left(\alpha+\frac{1}{8}\right)}>0\;,
\end{equation}
since 
here
the quadratic function of
 $V_*^2$ was shown in the proof of
  Proposition \ref{propd1} to be positive for $V_*^2\le C_R^2M/\mu_2$.

 \begin{prop}
\label{L0positive}
For $\alpha>0$, $V_*^2\le C_R^2m/\mu_2$ the constant $L_0$ in $(\ref{L0})$ is positive.
\end{prop}

\emph{Proof. }
\emph{Part 1}. Consider the cases when $a)$ $\alpha\ge 1$, $0< V_*^2\le C_R^2m/\mu_2$ or $b)$ $-1/2+\sqrt{2}\le \alpha\le 1$ so that  $\frac{3}{2}[\alpha-1+\sqrt{1-\alpha}]\le V_*^2\le C_R^2m/\mu_2$.

For the parameter values in both $a)$ and $b)$,  $M_0$ defined in (\ref{M0}) is nonnegative.

We show that $B_1$ is negative, which implies $B_2$ is also negative. The inequality $B_1<0$ gives
\begin{equation}\label{B1eq1}
{2V_*^4-3(\alpha+1)V_*^2+\frac{9}{4}\alpha}{}> V_*^2M_0\;.
\end{equation}
Here, this is valid for 
\[V_*^2< \frac{3\alpha+3-3\sqrt{\alpha^2+1}}{4}\;, \] 
which is the case, since 
\[\frac{C_R^2m}{\mu_2}=\frac{2\alpha+1-\sqrt{3\alpha^2+(\alpha-1)^2}}{4}< \frac{3\alpha+3-3\sqrt{\alpha^2+1}}{4}\quad \text{ for }\alpha>0\;. \]
Indeed, the above provides us with
\begin{equation}\label{eqimpineq}
3\sqrt{\alpha^2+1}-\sqrt{3\alpha^2+(\alpha-1)^2}< \alpha+2\;,
\end{equation}
and the left-hand side is positive as a result of 
\[9(\alpha^2+1)> 3\alpha^2+(\alpha-1)^2\;. \]
Thus from (\ref{eqimpineq}) we have
\[(\alpha^2+1)( 3\alpha^2+(\alpha-1)^2) -(2\alpha^2-\alpha+1) =2\alpha^3>0\;, \quad \text{ for }\quad \alpha>0\;.\]
Then (\ref{B1eq1}) leads to 
\[\frac{3}{16} (8V_*^2-3)(4V_*^2-3\alpha)^2<0\;,\] 
if $V_*^2 < 3/8$,
and this holds by Proposition \ref{prog}. Therefore $B_j <0$, for $j=1,2$. This together with (\ref{pos1}), (\ref{pos2}) and (\ref{pos3}) shows $L_0>0$.\hfill $\Box$

\vspace{0.1in}\emph{Part 2.} Consider the cases when $a)$ $0<\alpha< -1/2+\sqrt{2}$, $0<V_*^2\le C_R^2m/\mu_2$ or $b)$ $-1/2+\sqrt{2}\le\alpha\le 1$, $0<V_*^2\le \frac{3}{2}[\alpha-1+\sqrt{1-\alpha}]$.


For the parameter values in cases $a)$ and $b)$,   $M_0$ in (\ref{M0}) is purely imaginary and can be written as
\begin{equation}\label{M0complex}
M_0={\rm i}R_0\;, \quad R_0=\sqrt{3(1-\alpha) \left(3\alpha -{4V^2_*}\right)-{4 V^4_*}}\;,
\end{equation}
where according to the proof of Proposition \ref{propd1}, $R_0^2>0$ since $V^2_* \le C_R^2m/\mu_2$.
Then
\[d_2=\overline{d_1} \quad \text{and }\quad B_2=\overline{B_1}\;,\]
and 
from (\ref{L0})
\[L_0=
\frac{|d_1| \text{Re}(\overline{B_1}\sqrt{d_1})}{\sqrt{3}
(3-8V_*^2)(2V_*^2-3\alpha)(V_*^4-\frac{1}{2}(2\alpha+1)V_*^2+\frac{3\alpha}{8})}\;.
\]
By  (\ref{pos1}), (\ref{pos2}) and (\ref{pos3}) it remains to show that $\text{Re}(\overline{B_1}\sqrt{d_1})$ is negative. We have
\begin{equation*}\label{Num1}
\text{Re}(\overline{B_1}\sqrt{d_1})={\rm Re}(B_1){\rm Re}(\sqrt{d_1})+{\rm Im}(B_1){\rm Im}(\sqrt{d_1})\;,
\end{equation*}
with
\begin{equation}\label{ReB1ImB1}\begin{array}{l}
\displaystyle{{\rm Re}(B_1)=-\frac{1}{3}V^4_*+\frac{\alpha+1}{2}V^2_*-\frac{3}{8}\alpha\;,\quad {\rm Im}(B_1)=\frac{1}{6}V_*^2R_0>0\;.}
\end{array}
\end{equation}
Note that due to (\ref{M0complex}) and (\ref{rad1}), $-\frac{\pi}{2}< \text{arg}(\sqrt{d_1})< 0$, { therefore }
\begin{equation}\label{ineqd1}
{\rm Re}(\sqrt{d_1})> 0\;,\quad {\rm Im}(\sqrt{d_1})< 0 \;.
\end{equation}
The roots of ${\rm Re}(B_1)$ are 
\begin{equation*}\label{rootsREB1}
v^2_\pm=\frac{3}{4}[\alpha+1\pm \sqrt{\alpha^2+1}]>0 \quad \text{ for }\alpha>0\;.
\end{equation*}
Considering case $b)$, we have
\[V_*^2<v^2_-\;, \quad \text{ for }\quad 3/4< \alpha\le 1\;,\]
and then  ${\rm Re}(B_1)< 0$. This with (\ref{ReB1ImB1}), (\ref{ineqd1}) implies 
$\text{Re}(\overline{B_1}\sqrt{d_1})$
is negative and part $b)$ is proved.

Returning to part $a)$, for ${\rm Re}(B_1)<0$ we require 
\[\frac{2\alpha+1-\sqrt{3\alpha^2+(\alpha-1)^2}}{4}<v_-^2\;,\]
which was shown to be valid for $\alpha>0$ in  the proof of  Proposition \ref{propd1}, section \ref{WHeqcrack}. 
Therefore, the proof 
 is complete. \hfill$\Box$

\renewcommand{\theequation}{II-\arabic{equation}}    
  \setcounter{equation}{0}  
\section*{Appendix II: Poles of $L(\xi)$}

Here, for $\mu_1=\mu_2=1$, we  verify  the  inequalities (\ref{eqreg1}) and (\ref{eqreg3}) of section \ref{RPLXI},
which indicate when the functions (\ref{Droot2}) and $(\ref{Droot3})$ are solutions of  (\ref{hdelt}), respectively.

 For $\xi \in [0, 4\pi]$,  $\sign\{r(\Lambda_1^*)/r(\Lambda_2^*)\}=1$ when $\xi^2V^2= (\Omega^{(D)}_2)^2$. 
By direct substitution of (\ref{Droot2}) into (\ref{hdelt}) we obtain
\begin{equation}\label{hdelt2}
q\Big|_{V^2=(\Omega^{(D)}_2)^2/\xi^2}=\pm \frac{3}{2}{\text{i} \sin^2(\xi/2) |\tan(\xi/2)||4\cos^2(\xi/2)-1|}\{
 T_{1}(\xi)-
T_{2}(\xi)\}\;,
\end{equation}
where 
\begin{eqnarray*}
T_j(\xi)&=&\text{sign}\{2(1+(-1)^j\text{sign}\{\cos(\xi/2)\})\cos^2(\xi/2)-1\}\label{d12}\;,\quad j=1,2
\end{eqnarray*}
and the sign in front of the right-hand side of (\ref{hdelt2}) depends on the sign of $r(\Lambda_1^*)$. 
The factor in front of the curly braces in (\ref{hdelt2}), is zero when $\xi=0, \frac{2}{3}\pi, \frac{4}{3}\pi,  2\pi, \frac{8}{3}\pi, \frac{10}{3}\pi$, and $4\pi$, and is singular at $\xi=\pi$ and $3\pi$. 

\begin{figure}[htbp]
\centering
\includegraphics[width=0.8\textwidth]{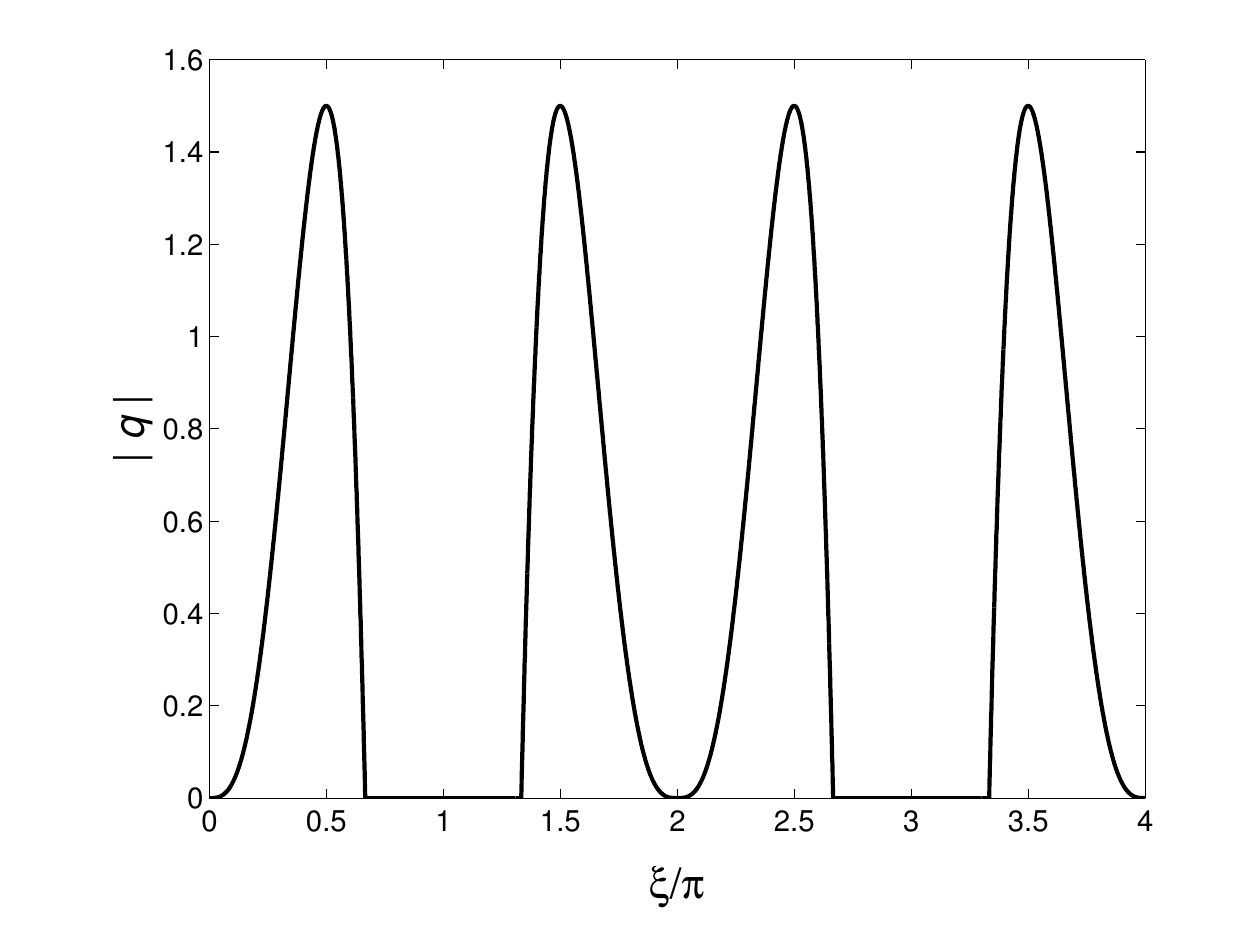}
\caption{
The plot of the absolute value of $q$ in (\ref{hdelt2}) as a function of the wave number, when $V^2\xi^2=(\Omega_2^{(D)})^2$. }
\label{App_1}
\end{figure}

We now show that (\ref{hdelt2}) is zero in the neighbourhood of these poles, by solving the equation
\[{T_1(\xi)}-{T_2(\xi)}=0\;.\]
For $\xi=\pi, 3\pi$ it is clear this equation is satisfied. When $\xi \ne \pi, 3\pi$, we are left to determine when the inequality
\[1-4\cos^2(\xi/2)>0\;,\]
is satisfied. This occurs for $\xi\in[\frac{2}{3}\pi, \frac{4}{3}\pi]\cup[\frac{8}{3}\pi, \frac{10}{3}\pi]$.  
 The above
  is 
  confirmed in Figure \ref{App_1}.
\begin{figure}[htbp]
\centering
\includegraphics[width=0.8\textwidth]{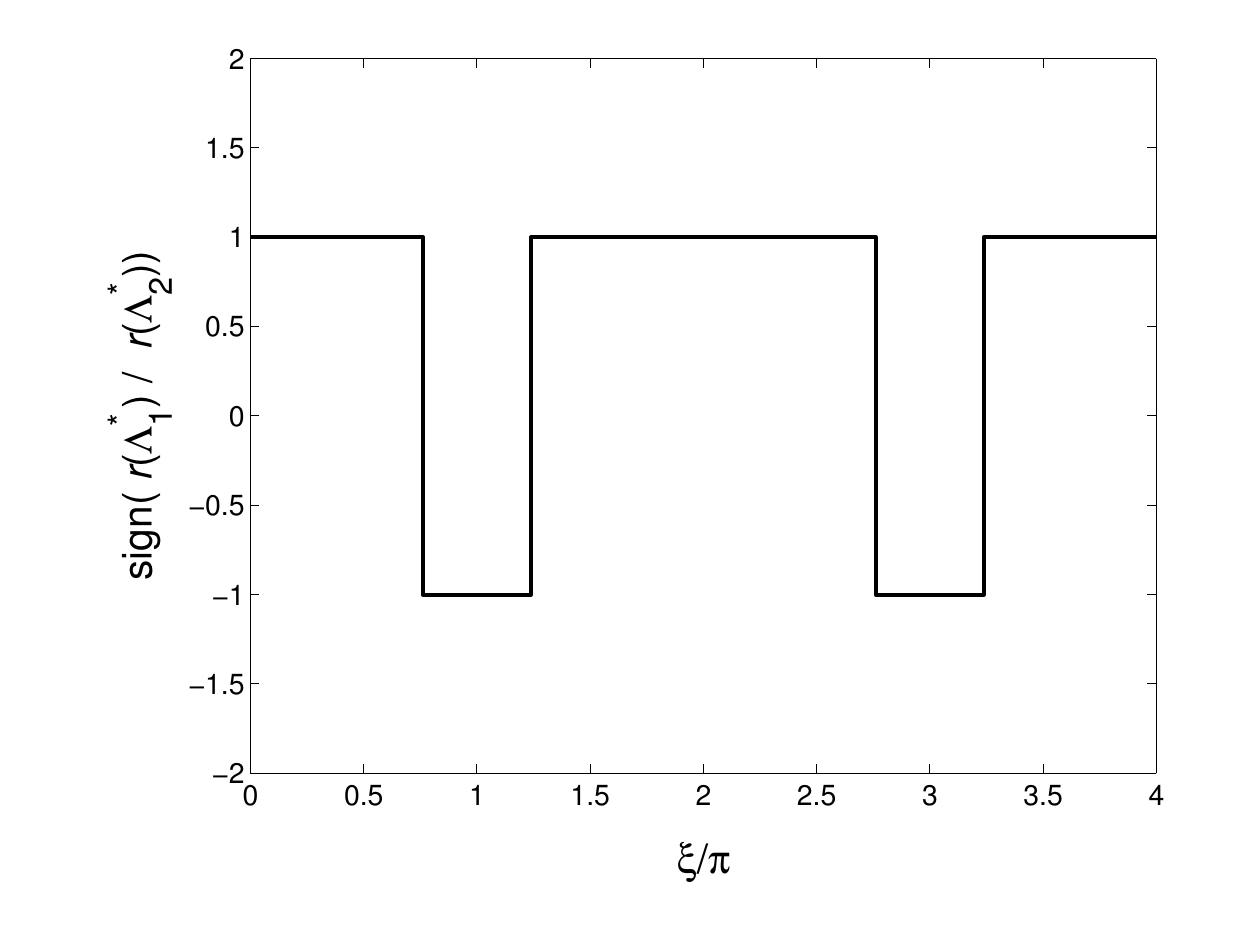}
\caption{The
sign of the ratio of $r(\Lambda_1^*)$ to $r(\Lambda_2^*)$ as a function of $\xi/\pi$, when $V^2\xi^2=(\Omega_3^{(D)})^2$. }
\label{App_2}
\end{figure}

  Next, insertion of  (\ref{Droot3}) into (\ref{hdelt}) yields
  \begin{eqnarray}
  q\Big|_{V^2=(\Omega^{(D)}_3)^2/\xi^2}=\frac{\text{i}\sqrt{3}\sin^2(\xi/2)|\sin(\xi/2)|(2\cos(\xi/2)-1+\sqrt{3})\sqrt{\sqrt{3}-2\sin^2(\xi/2)}}{2\sqrt{2\cos^2(\xi/2)+\sqrt{3}}}\{r(\Lambda_1^*)
  +r(\Lambda_2^*)
  \}\;.\label{hdelt3}
  \end{eqnarray}
The factor outside the curly braces in  (\ref{hdelt3}) has the roots $\xi=0, 0.7614\pi,1.2386\pi,2\pi,2.7614\pi, 3.283\pi$ and  $4\pi$. 
We must have $\sign\{r(\Lambda_1^*)/r(\Lambda_2^*)\}=-1$ when $V^2\xi^2=(\Omega_3^{(D)})^2$. Figure \ref{App_2}, shows the plot of $\sign\{r(\Lambda_1^*)/r(\Lambda_2^*)\}$ in this case, and we see that this condition holds provided $\xi$ satisfies the inequalities of (\ref{eqreg3}).

 \renewcommand{\theequation}{III-\arabic{equation}}    
  \setcounter{equation}{0}  
 \section*{Appendix III: Asymptotic properties of $L(\xi)$}

We now state and prove some properties of the kernel function $L(\xi)$, which are 
used in the analysis of  
the Wiener-Hopf equation (\ref{eqWH}),  in  section 6.
\begin{lem}\label{lemLinf}
$L(\xi)\to 1$ as $|\xi|\to \infty$. 
\end{lem}

\emph{Proof. }
According to (\ref{eqlam1a}),  for $|\xi|\to  \infty$, we have
\[\Lambda_j^*=\frac{1}{2z_j}+O(|z_j|^{-3})\;, \quad \text{ and }\quad r(\Lambda_j^*)=1\;, j=1,2\;,\]
\begin{equation}\label{asqnj}
\sqrt{z_j^2-1}=z_j+O({|z_j|}^{-1})\;,
\end{equation}
\[F(z_j)=3z_j^2-\frac{m}{\mu_2}(\varepsilon+{\rm i} \xi V)^2z_j[ 2 \cos(\xi/2)+1]+O(|\xi|^2)\;.\]
Since $z_j=O(|\xi|^2)$ for $|\xi|\to \infty$, then
\[F(z_2)\sqrt{z_1^2-1}-F(z_1)\sqrt{z_2^2-1}=3z_1z_2(z_2-z_1)+O(|\xi|^4)\;,\]
\[3(z_2-z_1)\sqrt{z_1^2-1}\sqrt{z_2^2-1}=3z_1z_2(z_2-z_1)+O(|\xi|^2)\;.\]
The leading order term in both of the above estimates is $O(|\xi|^6)$ for $|\xi|\to  \infty$, therefore 
\[L(\xi)=1+O(|\xi|^{-2}), \quad |\xi|\to  \infty\;.\]\hfill $\Box$

Next we state
\begin{lem} \label{lem3} For the product of two complex functions $f(\xi)$, $g(\xi)\in\CJ$, we have $(fg)(\xi)\in \CJ$.   
\end{lem}
This leads to 
\begin{lem}\label{lem2} For $\xi \in \mathbb{R}$, for $\varepsilon> 0$, $L(\xi) \in \CJ$.
\end{lem}

\emph{Proof.}
 The term
\[\frac{m}{\mu_2}(\varepsilon^2-\xi^2V^2+2{\rm i} \xi \varepsilon V)\in \CJ\]
since it has an even real part and odd imaginary part with respect to $\xi$.
This together with Lemma \ref{lem3} and (\ref{eqzj}) shows that $z_j\in \CJ$, $j=1,2$. 
and hence, after a second application of    
Lemma \ref{lem3}  with (\ref{Lkn1n2}) 
we complete the proof. 

Now we consider the index of the function $L(\xi)$ which defined by 
\begin{equation}\label{eqInd}
{\rm Ind}(L(\xi))=\frac{1}{2\pi} [{\rm arg}(L(\infty))-{\rm arg}(L(-\infty))]\;.
\end{equation}
This quantity can also be interpreted as the number of times the path traced by $L(\xi)$ in the complex plane for $\xi \in (-\infty,  \infty)$ winds around the origin. We have

\begin{lem}\label{lemind}
For $\varepsilon> 0$, we have ${\rm Ind}(L(\xi))=0$.
\end{lem}

\emph{Proof.}
According to section 6, if $\xi=\xi^*$ represents a root or a pole of $L(\xi)$ such that $V<V_g$ $(V>V_g)$, then the regularisation parameter $\varepsilon$ shifts this root or pole into lower (upper) half of the complex plane. Correspondingly, $\xi=-\xi^*$ is then a root or a pole of $L(\xi)$ such that $V>V_g$ ($V<V_g$) and this will be located in the upper (lower) half of the complex plane after the regularisation of $L(\xi)$. Therefore, $L(\varepsilon+{\rm i}\xi V,\xi )$ has an equal number of roots  in the upper and lower half planes 
 and an equal number of poles  in the upper and lower half planes, 
 and these numbers are independent of $\varepsilon$. Hence it is sufficient to show that for large $\varepsilon$, ${\rm Ind}(L(\varepsilon+{\rm i}\xi V, \xi))=0$.

First assume $\xi, \varepsilon\gg 1$, so that $| \varepsilon+{\rm i }\xi V |\gg1$, then 
\[z_j=O(\varepsilon^2+\xi^2V^2)\;,\qquad \Lambda_j^*=\frac{1}{2z_j}+O\Big(\frac{1}{(\varepsilon^2+\xi^2V^2)^3}\Big)\;,\]
and so 
\[r(\Lambda_j^*)=1\;,\qquad 3(z_2-z_1)\sqrt{z_1^2-1}\sqrt{z_2^2-1}=3z_1z_2(z_2-z_1)+O(\varepsilon^2+\xi^2V^2)\;,\]
and 
\[\begin{array}{c}
 F(z_2)\sqrt{z_1^2-1}-F(z_1)\sqrt{z_2^2-1}\\ \\
= \displaystyle{3z_1z_2(z_2-z_1)-\frac{m}{\mu_2}(\varepsilon+{\rm i} \xi V)^2(z_2-z_1)(2+\cos(\xi/2))+O(\varepsilon^2+\xi^2V^2)\;.}
\end{array}\]
Thus, from (\ref{Lkn1n2}), 
\begin{equation}\label{LOP}
L(\varepsilon+{\rm i}\xi V, \xi)=1+\frac{m(\varepsilon+{\rm i}\xi V)^2[2+\cos(\xi/2)]}{3\mu_2 z_1z_2}+O\Big(\frac{1}{(\varepsilon^2+\xi^2V^2)^2}\Big)\;,
\quad \varepsilon^2+\xi^2V^2 \to \infty\;,
\end{equation}
where according to (\ref{eqzj}), 
$z_1z_2$ is $O((\varepsilon^2+\xi^2V^2)^2)$.
 The remainder estimate here is uniform for large  $\varepsilon$ and $\xi$.

\begin{figure}[htbp]
\centering
\includegraphics[width=0.9\textwidth]{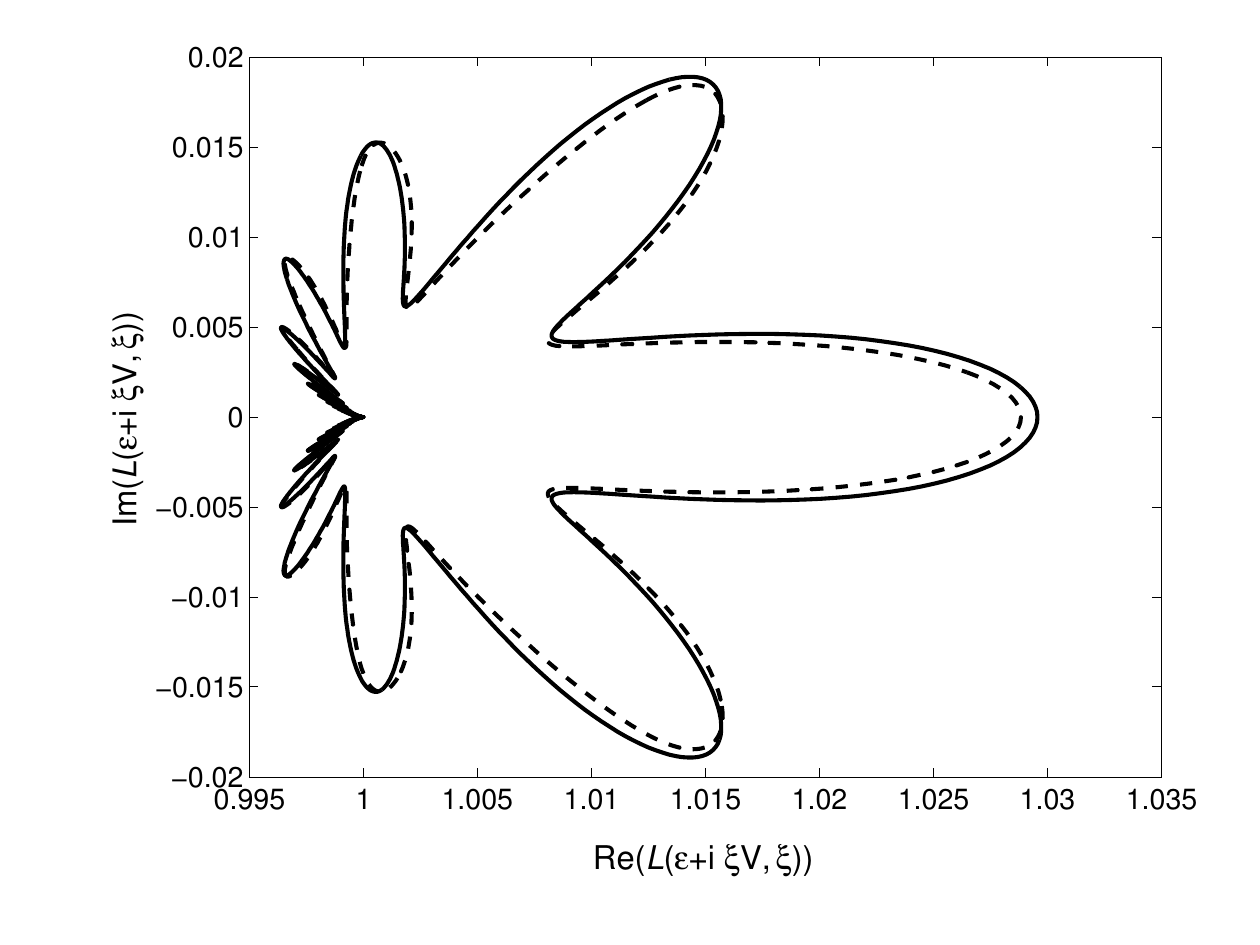}
\caption{The contour traced by $L(\varepsilon+{\rm i}\xi V,\xi)$ in the complex plane for $\varepsilon=10$,  $m=\mu_1=\mu_2=1$, $(\alpha=1)$, $V=0.3941$, $\xi /\pi\in (-400,400)$. The computations for the solid curve are based on formula (\ref{Lkn1n2}) and those for the dashed curve are based on the leading order part of  (\ref{LOP}).}
\label{FigLOP}
\end{figure}

The plot of 
the contour traced by $L(\xi)$ for large $\varepsilon$ and $\xi$, with $m=\mu_1= \mu_2=1$ $(\alpha=1)$, $V=0.3941$,  
 can be found in Figure \ref{FigLOP}. In this figure both the plot given by (\ref{Lkn1n2}) and the leading order term of (\ref{LOP}) are presented, and it can be seen there is a good agreement between both computations.   
 When $\varepsilon \gg \xi$, from (\ref{LOP})
 \[L(s+{\rm i}\xi V, \xi)=1+\frac{\mu_2(2+\cos(\xi/2))}{m}\left[\frac{1}{\varepsilon^2}-\frac{2{\rm i}\xi V}{\varepsilon^3}\right]+O\Big(\frac{1}{\varepsilon^4}\Big)\;,\quad \varepsilon\to \infty\;,\]
and this asymptotic expression (\ref{LOP}) implies that for large $\varepsilon$ the contour traced by $L$ does not cross the negative real axis, since to leading order the above real part is always positive, whereas the imaginary part can  cross the real axis for $\xi \in(-\infty, \infty)$. Similar behaviour can be seen in  Figure \ref{FigLOP}, where the contour traced by $L(\varepsilon+{\rm i }\xi V, \xi)$ for $\xi \in (-400\pi, 400\pi)$ in the complex plane forms a closed loop which passes through 1. Therefore, Ind$(L(\varepsilon+{\rm i}\xi V, \xi))=0$ in the limit $\varepsilon \to \infty$ and this also holds for all $\varepsilon> 0$.\hfill $\Box$

\vspace{0.1in}The next two corollaries provide us with additional properties of $L(\xi)$ needed for the factorisation of (\ref{eqWH}).
As a corollary of Lemma \ref{lem2}
 we have
 \begin{coro}\label{coroargodd}For $\xi \in \mathbb{R}$, $\varepsilon>0$,
\[{\rm arg }(L(\xi))=-{\rm arg }(L(-\xi))\;.\]
\end{coro}

Owing to   (\ref{eqInd}), Lemma \ref{lemind}
 and Corollary \ref{coroargodd} we also have: 
 
\begin{coro}\label{coroarg0}
${\rm arg }(L (\xi))=0$ as $\xi \to \pm \infty$.
\end{coro}


\renewcommand{\theequation}{IV-\arabic{equation}}    
  \setcounter{equation}{0}  
\section*{Appendix IV: Solution of the Wiener-Hopf equation (\ref{eqfactored}) for a particular load}
In order to determine the functions $\mathcal{S}_+$ and $\mathcal{S}_-$, as solutions of the Wiener-Hopf equation (\ref{eqfactored}), for the sake of simplicity, we consider a certain type of ``load''  $\phi^F$ which produces simple examples for the additive split $P_++P_-$ 
in the right-hand side of  (\ref{eqfactored}).

Let $\cP_{+}$ ($\CP_-$) be the set of poles  and $\CZ_+$ $(\CZ_-)$ the set of zeros of $L(\varepsilon+{\rm i}\xi V, \xi)$ that are located in the upper (lower) half plane. 
The first factor in the right-hand side of (\ref{eqfactored}) is singular at the zeros of $L_-$ and the singular points of $L_+$. 
Assume  $\xi^-_p\in \CP_-$, and
\begin{equation}\label{Lplusasymp1}
L_+(\varepsilon+{\rm i}\xi V, \xi)\sim \frac{A_p}{[ \zeta_- \varepsilon-{\rm i}(\xi-\xi^-_p) ]^{a}}\;,\quad a>0, \varepsilon\to +0\;,
\end{equation}
where $\zeta_-=1/(V_g-V)$ for $\xi \to \xi^-_p$. Also, suppose for $\xi_z^+\in \CZ_+$
\begin{equation}\label{Lplusasymp}
L_-(\varepsilon+{\rm i}\xi V, \xi)\sim {A_z}{[ \zeta_+ \varepsilon-{\rm i}(\xi-\xi_z^+) ]^{b}}\;,\quad b >0, \varepsilon \to +0\;,
\end{equation}
for $\xi \to \xi_z^+$ and $\zeta_+=1/(V-V_g)$. Here, $\xi_p^-$ is a pole and $\xi_z^+$ is a zero of $L$ located in the lower and upper  half of the complex plane, respectively, and the constants $A_p$ and $A_z$ depend on $\alpha$.

We consider two examples  of the function $\phi$: 

a) 
\begin{equation}\label{CaseAload}
\phi_A(\eta)=\phi(\eta)=-{\mathcal{C}_A(2\zeta_- \varepsilon)^a \text{exp}[(\zeta_- \varepsilon-{\rm i}\xi_p^-)\eta]H(-\eta)} \;,
\end{equation}
and after taking the Fourier transform, we have
\begin{equation}\label{qbc1}\phi_{A,-}=-\frac{(2\zeta_-\varepsilon)^a\CC_A}{[\zeta_-\varepsilon+{\rm i}(\xi-\xi_p^-) ]}\;, \quad \phi_{A,+}=0\;,\end{equation}
and
b) 
\[\phi_B(\eta)=q(\eta)={\mathcal{C}_B(2\zeta_+\varepsilon)^b \text{exp}[-(\zeta_+ \varepsilon+{\rm i}\xi_z^+)\eta]H(\eta)} \;.\]
so that 
\begin{equation}\label{qbc2}\phi_{B,+}=\frac{(2\zeta_+\varepsilon)^b\CC_B}{[\zeta_+\varepsilon-{\rm i}(\xi-\xi_z^+) ]}\;,\quad  \phi_{B,-}=0\;.\end{equation}

In cases a) and b), we have $\CC_A$  and $\CC_B$ are constants representing the intensity of the load. Also 
\begin{equation}\label{lims0}
\lim_{\varepsilon \to +0}\phi_A(\eta)=\lim_{\varepsilon \to +0}\phi_B(\eta)=0\;.
\end{equation}

\emph{Right-hand side of $(\ref{eqfactored})$ for the load of case $a)$.} We insert (\ref{qbc1}) into (\ref{eqfactored}) and taking the limit as $\varepsilon\to+0$ and noting that due to (\ref{lims0})
\[ \lim_{\varepsilon \to +0}\frac{\phi_{A,-}}{L_-}=0\;,\]
we have using (\ref{Lplusasymp1}):
\[L_+ \mathcal{S}_++\frac{\mathcal{S}_-}{L_-}=A_p\mathcal{C}_A\lim_{\varepsilon \to+0}\frac{ (2\zeta_-\varepsilon )^a}{[ \zeta_- \varepsilon -{\rm i}(\xi-\xi_p^-) ]^{a}[\zeta_-\varepsilon +{\rm i}(\xi-\xi_p^-) ]}\;.\]
Here, according to \cite[Section 2.2.4]{Slepyan}, the above limit converges to the Dirac delta function, 
 therefore
\begin{eqnarray*}
L_+ \mathcal{S}_++\frac{\mathcal{S}_-}{L_-}&\sim& 2\pi A_p\mathcal{C}_A\delta(\xi-\xi^-_p)= A_p\mathcal{C}_A\left[\frac{1}{\varepsilon+{\rm i}(\xi-\xi^-_p)} +\frac{1}{\varepsilon-{\rm i}(\xi-\xi^-_p)} \right]\;, \quad \varepsilon \to +0\;.
\end{eqnarray*}

\emph{Right-hand side of $(\ref{eqfactored})$ for the load of case $b)$.} Similarly, using (\ref{Lplusasymp}), (\ref{qbc2}) and 
\[ \lim_{\varepsilon \to +0}{\phi_{B,+}L_+}=0\;,\]
by \cite[Section 2.2.4]{Slepyan}, we have 
\begin{eqnarray*}
L_+ \mathcal{S}_++\frac{\mathcal{S}_-}{L_-}&=&\lim_{\varepsilon \to+0}\frac{\mathcal{C}_B}{A_z}\frac{ (2\zeta_+\varepsilon )^b}{[ \zeta_+ \varepsilon+{\rm i}(\xi -\xi^+_z) ]^{b}[\zeta_+\varepsilon -{\rm i}(\xi-\xi^+_z) ]}\\ \\
&\sim& 2\pi \frac{\mathcal{C}_B}{A_z}\delta(\xi-\xi_z^+)=\frac{\mathcal{C}_B}{A_z}\left[\frac{1}{\varepsilon+{\rm i}(\xi-\xi_z^+)} +\frac{1}{\varepsilon-{\rm i}(\xi-\xi_z^+)} \right]\;, \quad \varepsilon \to 0\;.
\end{eqnarray*}

\emph{General solution of $(\ref{eqfactored})$.} Consulting the above cases a) and b), by linear superposition  the right-hand side of   (\ref{eqfactored}) has the form
\begin{eqnarray*}\label{eqLgen}
L_+ \mathcal{S}_++\frac{\mathcal{S}_-}{L_-}&=& \sum_{\xi_p\in \CP_-} \CA_p\left[\frac{1}{\varepsilon+{\rm i}(\xi-\xi_p)} +\frac{1}{\varepsilon-{\rm i}(\xi-\xi_p)} \right]+\sum_{\xi_z\in \CZ_+} \CB_z\left[\frac{1}{\varepsilon+{\rm i}(\xi-\xi_z)} +\frac{1}{\varepsilon-{\rm i}(\xi-\xi_z)} \right]\;,
\end{eqnarray*}
for $\varepsilon \to 0\;,$ where $\CA_p$ and $\CB_p$ are arbitrary complex constants.
Then the functions $\mathcal{S}_+$ and $\mathcal{S}_-$ are given by
\begin{equation}\label{Qpm}
\begin{array}{c}
\displaystyle{\mathcal{S}_+=\frac{1}{L_+(\xi)} \left[\sum_{\xi_p\in \CP_-} \frac{\CA_p}{\varepsilon-{\rm i}(\xi-\xi_p)}+\sum_{\xi_z\in \CZ_+} \frac{\CB_z}{\varepsilon-{\rm i}(\xi-\xi_z)}\right] \;,}\\ \\
\displaystyle{\mathcal{S}_-=  L_-(\xi) \left[\sum_{\xi_p\in \CP_-} \frac{\CA_p}{\varepsilon+{\rm i}(\xi-\xi_p)}+\sum_{\xi_z\in \CZ_+} \frac{\CB_z}{\varepsilon+{\rm i}(\xi-\xi_z)}\right]\;,}
\end{array}
\end{equation}
where $\varepsilon \to +0$.
In section \ref{secWHE}, (\ref{Qpma})  follows by  considering terms in (\ref{Qpm})  which correspond  $p_-=0$ and $\nu=1/2$ in (\ref{Lplusasymp}) and (\ref{CaseAload}). 

\renewcommand{\theequation}{V-\arabic{equation}}    
  \setcounter{equation}{0}  
 \section*{Appendix V: Evaluation of  $L_\pm$ for $\xi \to 0$}

Now, using the Cauchy-type integral of (\ref{Cauchyint}) and the asymptotic representation (\ref{eqLasy}) for $L$ when $\xi \to 0$, we derive asymptotes of the functions $L_\pm$ near zero, found in
(\ref{asyLpmk0}).
The logarithmic term in (\ref{Cauchyint}) is rewritten as
\[{\rm ln }  \,L(\varepsilon+{\rm i}\xi V, \xi)={\rm ln} |L(\varepsilon+{\rm i}\xi V, \xi)|+{\rm i }\, { \rm arg}( L(\varepsilon+{\rm i}\xi V, \xi))\;.\]
Then 
\begin{eqnarray}\nonumber
L_\pm(\varepsilon+{\rm i}\xi V, \xi)&=&\text{exp}\Big(\pm\frac{1}{2\pi \text{i}} \int^\infty_{-\infty}\frac{\ln| L(\varepsilon+\text{i}s V, s)|}{s-\xi} d s \Big)\\ \nonumber\\
&&\times \text{exp}\Big(\pm\frac{1}{2\pi } \int^\infty_{-\infty}\frac{{\rm arg}( L(\varepsilon+\text{i}s V, s))}{s-\xi} d s\Big)\;.\label{derLpm1}
\end{eqnarray}
The first exponent on the right, by the Cauchy theorem, defines the modulus of $L(\varepsilon+{\rm i}\xi V, \xi )$, whereas the second exponent, owing to the expansion 
\[\frac{1}{s-\xi}=\sum_{n=0}^\infty \frac{\xi^n}{s^{n+1}}\]
yields 
\[\text{exp}\Big(\pm\frac{1}{2\pi } \int^\infty_{-\infty}\frac{{\rm arg}( L(\varepsilon+\text{i} s V, s))}{s-\xi} d s\Big)\sim \text{exp}\Big(\pm\frac{1}{2\pi } \int^\infty_{-\infty}\frac{{\rm arg}( L(\varepsilon+\text{i}s V, s))}{s} d s\Big)\;.\]
The integrand in the right-hand side is an even function as a result of Corollary \ref{coroargodd}, and so
\[\text{exp}\Big(\pm\frac{1}{2\pi } \int^\infty_{-\infty}\frac{{\rm arg}( L(\varepsilon+\text{i}s V, s))}{s-\xi} d s\Big)\sim \text{exp}\Big(\pm\frac{1}{\pi } \int^\infty_{0}\frac{{\rm arg}( L(\varepsilon+\text{i}s V, s))}{s} d s\Big)\;.\]
Therefore, allowing $\varepsilon \to +0$ in (\ref{derLpm1}),  and using Proposition \ref{Propasy0}, we obtain (\ref{asyLpmk0}).


\end{document}